\documentclass[lettersize,journal]{IEEEtran}
\usepackage{amsfonts}
\usepackage{algorithmic}
\usepackage{algorithm}
\usepackage{array}
\usepackage{textcomp}
\usepackage{stfloats}
\usepackage{url}
\usepackage{verbatim}
\usepackage{graphicx}
\usepackage{cite}
\usepackage{amsmath,bm}
\usepackage{cleveref}
\usepackage{amsthm}

\usepackage{bbm}
\usepackage{dsfont}
\usepackage{xcolor}

\newtheorem{theorem}{Theorem}
\newtheorem{lemma}{Lemma}
\newtheorem{proposition}{Proposition}

\newtheorem{definition}{Definition}

\usepackage{subfigure} 
\usepackage{balance}

\begin{document}

\title{Competitive Multi-armed Bandit Games for Resource Sharing}

\author{Hongbo Li,~\IEEEmembership{Member,~IEEE} and Lingjie Duan,~\IEEEmembership{Senior Member,~IEEE}

\thanks{Hongbo Li is with the Department of Electrical and Computer Engineering at The Ohio State University, Columbus, OH 43210, USA; Lingjie Duan is with the Pillar of Engineering Systems and Design, Singapore University of Technology and Design, Singapore 487372 (e-mail: li.15242@osu.edu, lingjie\_duan@sutd.edu.sg).}
\thanks{This work is also supported in part by the Ministry of Education, Singapore, under its Academic Research Fund Tier 2 Grant with Award no. MOE-T2EP20121-0001; in part by SUTD Kickstarter Initiative (SKI) Grant with no. SKI 2021\_04\_07; and in part by the Joint SMU-SUTD Grant with no. 22-LKCSB-SMU-053.}
}

\markboth{IEEE Transactions on Mobile Computing}%
{Shell \MakeLowercase{\textit{et al.}}: A Sample Article Using IEEEtran.cls for IEEE Journals}

\IEEEpubid{0000--0000/00\$00.00~\copyright~2021 IEEE}

\maketitle

\begin{abstract}
In modern resource-sharing systems, multiple agents access limited resources with unknown stochastic conditions to perform tasks.
When multiple agents access the same resource (arm) simultaneously, they compete for successful usage, leading to contention and reduced rewards.
This motivates our theoretical study of competitive multi-armed bandit (CMAB) games.
In this paper, we study a new $N$-player $K$-arm competitive MAB game, where non-myopic players (agents) compete with each other to form diverse private estimations of unknown arms over time. 
Their possible collisions on the same arms and the time-varying nature of arm rewards make the policy analysis here more involved than the existing studies for myopic players.
We explicitly analyze the threshold-based structures of the social optimum and the existing selfish policy, showing that the latter causes prolonged convergence times \(\Omega\big(\frac{K}{\eta^2}\ln({\frac{KN}{\delta}})\big)\), while the socially optimal policy with coordinated communication reduces it to \(\mathcal{O}(\frac{K}{N\eta^2}\ln{(\frac{K}{\delta})})\).
Based on the policy comparison, we prove that the competition among selfish players for the best arm can result in an infinite price of anarchy (PoA), indicating an arbitrarily large efficiency loss compared to the social optimum. 
We further prove that no informational (non-monetary) mechanism (including Bayesian persuasion) can reduce the infinite PoA, as strategic misreporting by non-myopic players undermines such approaches.
To address this, we propose a Combined Informational and Side-Payment (CISP) mechanism, which provides socially optimal arm recommendations with proper informational and monetary incentives to players according to their diverse and time-varying private beliefs. Our CISP mechanism keeps ex-post budget balanced for the social planner and ensures truthful reporting from players, thereby achieving the minimum \(\text{PoA}=1\) and the same convergence time as the social optimum.
\end{abstract}


\section{INTRODUCTION}
In today's resource-sharing systems, multiple agents access limited resources with unknown stochastic conditions to perform tasks.
For example, in wireless networks, multiple radio nodes opportunistically access a set of spectrum channels to transmit data to a base station or access point. 
For lightweight edge nodes like IoT devices, the quality of channel conditions is unknown and changes stochastically, necessitating them to learn different channels. 
These nodes often need to make decisions without expensive communication or coordination (which could also lead to privacy leakage) among them. 
Additionally, when multiple nodes access the same channel, they incur strong interference and compete with each other for successful transmission (\!\!\cite{li2020multi,jiang2025multi,liu2025optimizing}). 
Such a setting calls for not only solving a distributed multi-armed bandit (MAB) problem, but the challenging one where players (nodes) are competitive with each other. Unlike classical distributed MAB models (e.g., \cite{liu2018information,yang2022distributed,zhu2023distributed}), here the rewards of arms (channels) obtained by competitive players are no longer exogenous but depend on players' arm choices to possibly collide, significantly complicating their decision-making process.


Motivated by resource-sharing applications like cognitive radio networks and other transportation applications (\!\!\cite{li2023congestion,li2024distributed,li2024human}), distributed MAB problems with potential collisions among players have garnered significant interest in the past decade \cite{liu2010distributed,rosenski2016multi,marden2014achieving,bistritz2018distributed,boursier2019sic,wang2020optimal,xiong2023decentralized}. 
In these works, multiple players simultaneously make arm decisions with no or rather limited communication.
To address this challenge, \cite{rosenski2016multi} introduces a communication-free algorithm for players to randomly select different idle arms to reduce collision probabilities after a certain exploration stage. 
Furthermore, in \cite{marden2014achieving} and \cite{bistritz2018distributed}, players iteratively switch arms based on their private empirical probabilities to identify the optimal arms and ultimately converge on a single choice. \cite{boursier2019sic} and \cite{wang2020optimal} allow players to observe the number of competitors selecting the same arm during collisions, boosting the overall social reward closer to what centralized algorithms achieve. 
However, all of these works assume that players prioritize altruism in maximizing the total reward, overlooking the potential for self-interest and selfish deviations from the proposed algorithms.

\IEEEpubidadjcol

In practical scenarios of distributed MAB problems (e.g., communication networks, transportation, and queueing), players often exhibit selfish behaviors and may be reluctant to explore certain arms for others. This self-interest can lead to significant efficiency losses within the system (\!\!\cite{bolton1999strategic,branzei2021multiplayer,sentenac2021decentralized,boursier2020selfish,huang2023near,xu2023competing}).
For instance, \cite{branzei2021multiplayer} evaluates that competitive players engage in less exploration than their cooperative counterparts over the long term, resulting in a greatly reduced total reward. 
\cite{sentenac2021decentralized} considers strategic, non-myopic players engaged in online learning within queueing systems. Each server serves as an arm, and the queue length experiences instability due to selfish server decisions.
\cite{xu2023competing} proposes to facilitate the convergence to the Nash Equilibrium (NE). However, the proposed algorithms primarily cater to players' self-interests without considering any mechanism design to incentivize exploration and remedy efficiency losses.

In the existing literature on incentivized exploration, there are roughly two types of mechanisms. One is the informational (non-monetary) mechanism (e.g., Bayesian persuasion), and the other is the monetary/pricing mechanism.  
Initially, \cite{kremer2014implementing} first exploited Bayesian persuasion to provide incentives for socially optimal arm choices, by leveraging the information asymmetry between the social planner and myopic players. 
This mechanism is then widely adopted in subsequent research on regulating MAB games \cite{papanastasiou2018crowdsourcing,mansour2022bayesian,simchowitz2023exploration} and routing/congestion games \cite{tavafoghi2017informational,das2017reducing,li2023congestion,li2024human}. 
In the absence of information control, monetary mechanisms explore the incentivization of players through direct payments (e.g., \cite{frazier2014incentivizing} and \cite{che2018recommender}). 
However, both informational and monetary mechanisms require the social planner to possess enough information about the system, which relies on the assumption that players are myopic or have truthfully reported their past reward observations. 
In contrast, our competitive MAB game studies a more sophisticated scenario, where non-myopic players have access to their private past observations and are non-myopic to strategically misreport these private observations to mislead the social planner and improve their own long-term benefits.

In this paper, we study the $N$-player CMAB games for resource sharing, where each player is selfish and non-myopic to maximize its own long-term reward. 
Let us pose two key questions, each followed by a discussion of the challenges: \emph{how to analyze non-myopic players' competitive learning policies in CMAB games} and \emph{how to design an efficient mechanism to regulate them to approach social optimum}. 
\begin{itemize}
    \item The first challenge is to \emph{theoretically analyze and compare selfish and socially optimal policies via price of anarchy (PoA) for non-myopic players.} 
    In CMAB games, players do not communicate but compete with each other to form diverse private estimations of unknown arms over time. Their possible collisions on the same arms and the time-varying nature of arm rewards make the policy analysis here more involved than that for myopic players \cite{kremer2014implementing,mansour2022bayesian,li2023congestion}. 
    We aim to solve explicit structures of the selfish policy and the socially optimal policy, and further analytically compare the two policies via the worst-case PoA analysis. 
    Even if some recent MAB works also consider non-myopic selfish players (e.g., \cite{boursier2020selfish,branzei2021multiplayer,sentenac2021decentralized,huang2023near}), they miss explicit analysis of PoA. 
    \item The second challenge pertains to the \emph{social planner's mechanism design in the presence of information misreporting by players}. In our CMAB games, the social planner needs to collect arm information from the involved players, who serve as the information sources over time. However, non-myopic players with diverse past observations themselves may strategically misreport these private observations to mislead the social planner and improve their own long-term benefits. 
    As such, informational mechanisms such as Bayesian persuasion and selected information hiding become ineffective, as they rely on the information truthfully reported by myopic players as sources \cite{kremer2014implementing,tavafoghi2017informational,mansour2022bayesian,li2023congestion}. One may use monetary incentives to replace informational mechanisms here, yet we cannot borrow existing monetary mechanisms (e.g., \cite{frazier2014incentivizing,che2018recommender}), as non-myopic players will misreport to lower the charges or even earn rewards from the social planner.

\end{itemize}

The key novelty of this paper and our main contributions are summarized as follows.
\begin{itemize}
    \item \emph{First competitive MAB game with non-myopic players to analyze and regulate:} To our best knowledge, this paper is the first to analyze and regulate an $N$-player CMAB games for resource sharing, where non-myopic players do not communicate but compete with each other to form diverse private estimations of unknown arms over time. Their possible collisions on same arms and the time-varying nature of arm rewards make the policy analysis here more involved than that for myopic players. 
    As multiple players choosing the same arm incurs collision, we aim to leverage information learning of different arms to counteract collisions among selfish players and improve the long-term social reward. 
    \item \emph{Threshold-based solutions of policies and PoA analysis for CMAB games:} 
    Non-myopic players under the selfish policy aim to maximize their own long-term rewards, while the socially optimal policy wants to maximize the long-term total reward for all players.
    Although analyzing the interactive arm decisions of non-myopic players is complex, we have explicitly solved both policies to be in threshold-based structures. We analyze that the selfish policy causes the convergence time \(\Omega\big(\frac{K}{\eta^2}\ln({\frac{KN}{\delta}})\big)\) to explore arms, while the social optimum reduces it to \(\mathcal{O}(\frac{K}{N\eta^2}\ln{(\frac{K}{\delta})})\). 
    Based on the policy comparison, we prove that the competition among selfish players for the best arm can result in an infinite price of anarchy (PoA), indicating an arbitrarily large efficiency loss compared to the socially optimal policy.
    \item \emph{Budget balanced Combined Informational and Side-Payment (CISP) mechanism design against information misreport:} We further prove that no informational (non-monetary) mechanism (e.g., Bayesian persuasion) can reduce the infinite PoA, as non-myopic players have diverse past observations and may strategically misreport their arm observations to improve long-term benefits. Alternatively, we propose a CISP mechanism, which provides socially optimal arm recommendations with proper informational and monetary incentives to players according to their diverse and time-varying private beliefs. Our CISP mechanism keeps ex-post budget balanced for the social planner and ensures truthful reporting from players, thereby achieving the minimum \(\text{PoA}=1\) and the same convergence time as the socially optimal policy.
\end{itemize}

The rest of our paper is organized as follows. 
In \Cref{section2}, we first introduce the system model of the CMAB games and formulate the optimization problems for selfish and socially optimal policies. 
Then in \Cref{section4}, we conduct theoretical analysis on both policies and derive the PoA of the selfish policy. 
Following that, \Cref{section5} introduces our CISP mechanism and analyzes its optimal PoA. 
Finally, we conduct experiments to verify our key results in \Cref{section_simulation} and conclude this paper in  \Cref{section6}. For ease of reading, we summarize all the key notations in Table \ref{notation_table}.

\section{SYSTEM MODEL AND PROBLEM FORMULATIONS}\label{section2}

We consider a resource-sharing network with $N$ agents (players) operating over $K$ resources in an infinite discrete time horizon $\mathbb{T}=\{1,2,\cdots \}$. The actual condition of a resource $k\in\mathbb{K}:=\{1,\cdots, K\}$ at time $t\in\mathbb{T}$ is denoted by $r_k(t)\in\{0,1\}$, which reflects the resource's usability when only one player is utilizing it. Here, $r_k(t)=1$ represents a good condition, providing a positive reward, while $r_k(t)=0$ indicates a bad condition without any reward.
As in the most existing MAB literature (e.g., \cite{rosenski2016multi,besson2018multi,krishnasamy2021learning,mansour2022bayesian}), $r_k(t)$ satisfies a Bernoulli process over time with an unknown mean value $\mu_k\in(0,1)$:
\begin{align}
    r_k(t)=\begin{cases}
        1, &\text{with probability } {\mu_k},\\
        0, &\text{with probability } 1-{\mu_k}.
    \end{cases}\label{r_k(t)}
\end{align}

In this resource-sharing network, a set of players $\mathbb{N}:=\{1,\cdots,N\}$ selfishly choose their best resources out of $K$ arms at any $t\in\mathbb{T}$, and players may interfere and compete with others for selecting the same resource for positive reward.
As the actual condition of each resource varies over time under the unknown mean reward $\mu_k$, players must strategically try certain resources to learn accurate $\mu_k$ values, enabling them to make better decisions and enhance their long-term rewards.

\begin{table}[!t]
\renewcommand{\arraystretch}{1.3}
\caption{Key notations and their meanings in the paper}
\label{notation_table}
\centering
\begin{tabular}{|c|m{0.33\textwidth}|}
\hline
\textbf{Notation} & \textbf{Meaning}\\
\hline
\hline
$\mathbb{T}$ & The number of user arrivals at time $t$.\\
\hline
$\mathbb{K}$ & The arm set.\\
\hline
$N$ & The number of players.\\
\hline
$r_k(t)$ & The actual reward reward of arm $k$ at time $t$.\\
\hline
$\mu_k$ & The actual mean reward of arm $k$.\\
\hline
$\eta$ & The minimum reward gap between any two arms.\\
\hline
$\mathbb{N}_k(t)$ & The set of players choosing arm $k$ at time $t$.\\
\hline
$\sigma_n(t)$ & The collision indicator of player $n$ at time $t$.\\
\hline
$\pi_n(t)$ & The arm decision of player $n$ at time $t$.\\
\hline
$r_k^n(t)$ & The binary reward received by player $n$ at time $t$.\\
\hline
$\tilde{\mu}_k^n(t)$ & Player $n$'s empirical mean reward of arm $k$.\\
\hline
$c_k^n(t)$ & The number of times player $n$ has pulled arm $k$ before time $t$.\\
\hline
$\tilde{\bm{\mu}}^n(t)$ & The set of empirical mean rewards $\tilde{\mu}_k^n(t)$ for player $n$.\\
\hline
$\tilde{r}_k^n(t)$ & Player $n$'s expected immediate reward of choosing arm $k$ at $t$. \\
\hline
$V(\cdot)$ & The long-term expected total reward function.\\
\hline
$\mathcal{T}(t)$ & The exploration threshold.\\
\hline
$\Delta \mu_{j,k}(t)$ & The maximum exploration benefit of switching from arm $k$ to arm $j$ at time $t$.\\
\hline
$\rho$ & Discount factor.\\
\hline
$\delta$ & A small constant.\\
\hline
$p_i(t)$ & Monetary payments/charge by our CISP mechanism.\\
\hline
\end{tabular}
\end{table}



This problem can be modeled by an $N$-player (agent) competitive MAB problem to explore and share $K$-armed bandits (resources), where $K> N$ (e.g., \cite{rosenski2016multi,besson2018multi,krishnasamy2021learning,mansour2022bayesian}).
In the following, we first introduce our reward learning model for this $N$-player $K$-arm competitive bandit game. Then we formulate the optimization problems for both selfish and socially optimal policies.
In this paper, for any two arms $i,j\in\mathbb{K}$, we assume $|\mu_i-\mu_j|>\eta$, where $\eta>0$ is an infinitesimal. 

\subsection{Competitive Reward Learning Model}\label{section2-1}

\begin{figure}[t]
    \centering
    \includegraphics[width=0.45\textwidth]{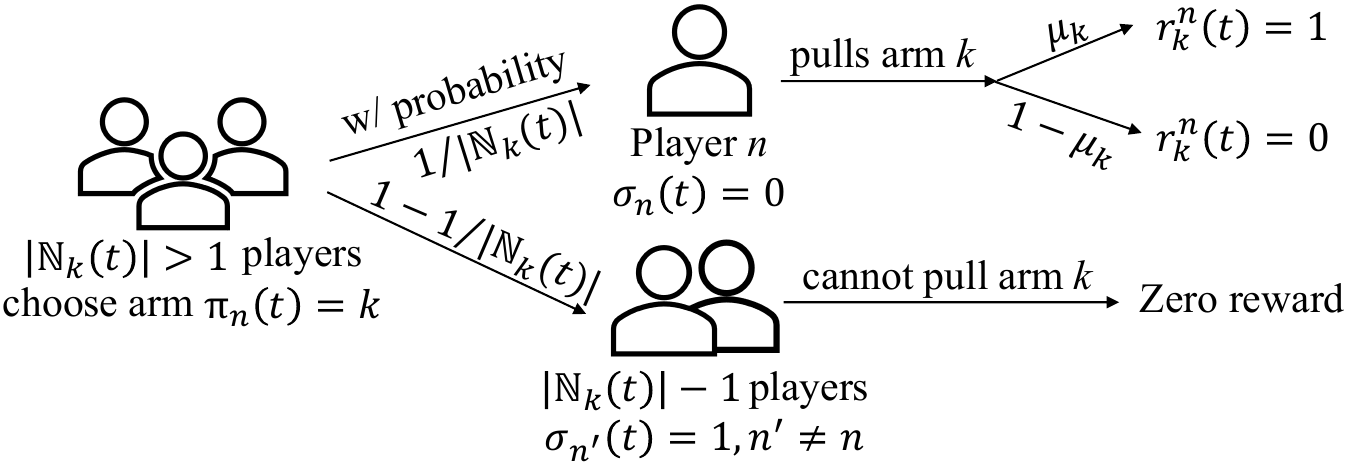}
    \caption{The process of arm pulling and reward observations when $|\mathbb{N}_k(t)|$ players simultaneously choose arm $k$. In this case, only one player is randomly selected with probability $\frac{1}{|\mathbb{N}_k(t)|}$ to pull arm $k$ (e.g., player~$n$ with $\sigma_n(t)=0$ in Fig.~\ref{fig:competitive_reward}) and receive a reward of $r_k^n(t)=r_k(t)$, where $r_k(t)$ is given in \cref{r_k(t)}.
    As in \cite{boursier2019sic,wang2020optimal}, the remaining $|\mathbb{N}_k(t)|-1$ players observe collisions involving $|\mathbb{N}_k(t)|$ players there and receive zero rewards (with $\sigma_{n'}(t)=1$). In other words, these $|\mathbb{N}_k(t)|-1$ players have no effective reward observation of this arm.}
    \label{fig:competitive_reward}
\end{figure}

In each time slot $t\in\mathbb{T}$, all competitive players in $\mathbb{N}$ make their arm decisions independently without any communication. 
Let $\pi_n(t)\in\mathbb{K}$ denote the arm decision of player~$n$ at time $t$, where $n\in\mathbb{N}$. Define $\sigma_n(t)\in\{0,1\}$ as the collision indicator for player $n$:
\begin{align*}
    \sigma_n(t)=\begin{cases}
        1, &\text{if player $n$ experiences a collision at time $t$},\\
        0, &\text{otherwise}.
    \end{cases}
\end{align*} 

\textbf{Reward model.} Define $\mathbb{N}_k(t):=\{n|n\in\mathbb{N},\pi_n(t)=k\}$ as the set of player(s) choosing arm $k$ at time $t$, and the number of players in this set is:
\begin{align}
    |\mathbb{N}_k(t)|=\sum_{n=1}^N \mathds{1}\{\pi_n(t)=k\}.\label{|N_k|}
\end{align}
After deciding on arm $\pi_n(t)=k$, let $r_k^n(t)\in\{0,1\}$ denote the binary reward received by player~$n$. Given the number of competitive players $|\mathbb{N}_k(t)|$ choosing the same arm $k$, this actual reward $r_k^n(t)$ falls into two cases:

1) $|\mathbb{N}_k(t)|=1$ without collision at arm $k$: Player $n$ pulls arm $k$ and receives reward $r_k^n(t)=r_k(t)$. 

2) $|\mathbb{N}_k(t)|>1$ with collisions at arm $k$: As shown in Fig.~\ref{fig:competitive_reward}, only one player is randomly selected with probability $\frac{1}{|\mathbb{N}_k(t)|}$ to pull arm $k$ (e.g., player $n$ with $\sigma_n(t)=0$ in Fig.~\ref{fig:competitive_reward}) and receive $r_k^n(t)=r_k(t)$. As in \cite{boursier2019sic,wang2020optimal}, the remaining $|\mathbb{N}_k(t)|-1$ players observe collisions involving $|\mathbb{N}_k(t)|$ players there and receive zero rewards (with $\sigma_{n'}(t)=1$). In other words, these $|\mathbb{N}_k(t)|-1$ players have no reward observation of this arm.
 
In summary, the expected immediate reward $r_k^n(t)$ received by any player $n\in\mathbb{N}_k(t)$ choosing arm $\pi_n(t)=k$ is:
\begin{align}
    \mathbb{E}[r_k^n(t)]=(1-\sigma_n(t))\mu_k,\label{r_k^n}
\end{align}
Since $\mu_k$ is unknown to players, it necessitates them to explore this arm to learn it. 

Next, we introduce how each player $n$ updates its empirical mean reward of arm $k$, denoted by $\Tilde{\mu}_k^n(t)$, based on its past arm decisions and corresponding reward observations.
Here we assume that players can observe collisions on the arm they pull at the current time.

\textbf{Diverse empirical mean rewards.} Initially, each player $n$ has a private preference for each arm $k$ with $\Tilde{\mu}_k^n(0)=\theta_k^n$. Then for any $t\in\mathbb{T}$, the update of empirical mean reward $\Tilde{\mu}_k^n(t)$ depends on player $n$'s past observed rewards and the number of times it has pulled arm~$k$. 

We define $c_k^n(t)$ as the number of times player~$n$ has pulled arm $k$ before time $t$. Then, following Fig.~\ref{fig:competitive_reward}, the update of $c_k^n(t+1)$ contains two possibilities. 
\begin{itemize}
    \item If player~$n$ chooses another arm $\pi_n(t)\neq k$, $c_k^n(t+1)$ remains unchanged.
    \item If player~$n$ chooses arm $\pi_n(t)=k$ and it is selected to pull this arm during time $t$, $c_k^n(t+1)$ increases by $1$ and it observes $r_k(t)$ of arm $k$ in \cref{r_k^n}). If it is not selected, $c_k^n(t+1)$ remains the same as $c_k^n(t)$ as there is no effective reward observation. 
\end{itemize}
According to the above two cases, we summarize the dynamics of $c_k^n(t+1)$ below:
\begin{align}
    c_k^n(t+1)=\begin{cases}
        c_k^n(t)+1, &\text{if $\pi_n(t)=k$ and $\sigma_n(t)=0$},\\
        c_k^n(t), &\text{otherwise.}
    \end{cases}\label{c_k^n}
\end{align}

Given player $n$'s historical reward $r_n^k(t)$ in \cref{r_k^n}) and successful pulling times $c_k^n(t+1)$ in \cref{c_k^n}), it updates its empirical mean reward $\Tilde{\mu}_k^n(t+1)$ of arm $k$ to:
\begin{align}
    \Tilde{\mu}_k^n(t+1)=\frac{\sum_{\tau=1}^t r_k^n(\tau)\mathds{1}\{\pi_n(\tau)=k\}}{c_k^n(t+1)},\label{tilde_mu}
\end{align}
where $\mathds{1}\{\pi_n(\tau)=k\}=1$ if $\pi_n(\tau)=k$ and $\mathds{1}\{\pi_n(\tau)=k\}=0$ otherwise. Note that $\Tilde{\mu}_k^n(t)=\theta_k^n$ if $c_k^n(t)=0$ without effective exploration. 

For ease of exposition, we summarize empirical mean reward $\Tilde{\mu}_k^n(t)$ of player $n$ in \cref{tilde_mu} among all arms into vectors: 
\begin{align*}
    \Tilde{\bm{\mu}}^n(t)=\big\{\Tilde{\mu}_k^n(t)|k\in\mathbb{K}\big\}.
\end{align*}
Based on $\Tilde{\bm{\mu}}^n(t)$, non-myopic player $n$ follows the selfish policy to always choose the arm that maximizes its long-term expected reward. 
While the socially optimal policy aims to maximize the total long-term reward for all players. Then we formulate the optimization problems for the two policies in the next subsection.

\subsection{Problem Formulations}
\textbf{Selfish policy.} First, we formulate the optimization problem for non-myopic players following the selfish arm-decision policy for its own benefit. The selfish policy is defined below.
\begin{definition}[Selfish policy]\label{def:selfish_policy}
Under selfish policy, each non-myopic player $n$ aims to maximize its own long-term $\rho$-discounted reward, where $\rho\in(0,1)$ is the discount factor.
\end{definition}

\begin{algorithm}[t]
\caption{Decision-making under the selfish policy}
\small
\label{algo:selfish}
\begin{algorithmic}[1]
\STATE \textbf{Input:} $\mathbb{T},\mathbb{K}$;\
\STATE Initialize $\Tilde{\mu}^n_k(0)=\theta_k^n,c_k^n(0)=0$, $\forall k\in\mathbb{K}$;\
\FOR{$t\in\mathbb{T}$}
\STATE Solve \cref{pi_n_s(t)} to obtain $\pi^{(s)}_n(t)$;\
\IF{$\sigma_{\pi^{(s)}_n(t)}(t)=0$}
\STATE Observe ${r}_k(t)$ and update $c_k^n(t+1)=c_k^n(t)+1$ by \cref{c_k^n};\
\STATE Update $\Tilde{\mu}^n_k(t+1)$ by \cref{tilde_mu};\
\ENDIF
\ENDFOR
\end{algorithmic}
\end{algorithm}

According to \Cref{def:selfish_policy}, each player needs to estimate its long-term reward for decision making.
Let $\tilde{r}_k^n(t)$ denote player $n$'s expected immediate reward of choosing arm $k\in \mathbb{K}$ at time $t$, which is estimated by player $n$ based on \cref{r_k^n} and \cref{tilde_mu}:
\begin{align}\label{E[r_k^n]}
    \tilde{r}_k^n(t)=\textbf{Pr}(\sigma_n(t)=0|\tilde{\bm{\mu}}^n(t))\cdot \tilde{\mu}_k^n(t),
\end{align}
where $\textbf{Pr}(\sigma_n(t)=0|\tilde{\bm{\mu}}^n(t))=\frac{1}{|\mathbb{N}_k(t)|}$ according to Fig.~\ref{fig:competitive_reward} and \cref{|N_k|}.
To simplify notations, we use $\textbf{Pr}(\sigma_n(t)=0)$ to denote $\textbf{Pr}(\sigma_n(t)=0|\tilde{\bm{\mu}}^n(t))$ in the following.
Since player $n$ is unaware of the exact number of players $|\mathbb{N}_k(t)|$ selecting arm~$k$ in \cref{E[r_k^n]} before making its arm decision, it estimates the expected number of players choosing arm $k$ based on its own empirical mean reward $\Tilde{\mu}_k^n(t)$. Next, we establish the following lemma to better justify how players make decisions based on $\Tilde{\mu}_k^n(t)$.
\begin{lemma}\label{lemma:equal_expect}
For player $n$ with empirical mean reward set $\tilde{\bm{\mu}}^n(t)$ at $t$, we have $\mathbb{E}[\tilde{\bm{\mu}}^n(t+1)]=\tilde{\bm{\mu}}^n(t)$.
\end{lemma}
The proof of Lemma~\ref{lemma:equal_expect} is given in Appendix~A.
The result implies that player $n$'s expected empirical mean reward remains unchanged in the next time slot $t+1$.
Consequently, player $n$ can reasonably believe that other players symmetrically have the same expected empirical mean reward $\tilde{\bm{\mu}}^n(t)$ from $t=0$, as this belief is the most statistically likely to hold. To support this assumption, we further validate it through simulations in Section~\ref{section_simulation}.
Based on this belief, players behave strategically, and their decisions collectively form a Nash equilibrium at time $t$. Consequently, player $n$ estimates $|\mathbb{N}_k(t)|$ at the current time $t$ by solving the Nash equilibrium under $\tilde{\bm{\mu}}^n(t)$, using established methods such as value iteration.

Let $\pi^{(s)}_n(t)\in\mathbb{K}$ denote the arm decision of player $n\in \mathbb{N}$ under the selfish policy, where the subscript $(s)$ tells the selfish arm-decision policy in Definition~\ref{def:selfish_policy}.
For ease of exposition, we summarize arm decisions of all players into vector $\bm{\pi}^{(s)}(t)=\{\pi^{(s)}_1(t),\cdots, \pi_N^{(s)}(t)\}$. 
Define $V^{(s)}_n(\tilde{\bm{\mu}}^n(t))$ as the long-term expected total reward of player $n$ under the selfish policy from \Cref{def:selfish_policy}.
At any time $t\in\mathbb{T}$, we leverage the Markov decision process (MDP) to formulate:
\begin{align}\label{pi_n_s(t)}
    V^{(s)}_n(\tilde{\bm{\mu}}^n(t))=\max_{\pi_n^{(s)}(t)\in\mathbb{K}} &\Big\{\tilde{r}_{\pi_n^{(s)}(t)}^n(t)\\&+\rho\mathbb{E}\Big[V^{(s)}_n(\tilde{\bm{\mu}}^n(t+1))\Big|r_{\pi_n^{(s)}(t)}^n(t)\Big]\Big\}\notag,
\end{align}
where $\tilde{r}_{\pi_n^{(s)}(t)}^n(t)$ is defined in \cref{E[r_k^n]}, and the cost-to-go term $\mathbb{E}[V^{(s)}_n(\tilde{\bm{\mu}}^n(t+1))|r_{\pi_n^{(s)}(t)}^n(t)]$ includes the following cases, based on the actual observed reward $r_k^n(t)$ to update $\tilde{\bm{\mu}}^n(t+1)$:
\begin{itemize}
    \item $\sigma_{\pi_n^{(s)}(t)|\tilde{\bm{\mu}}^n(t)}(t)=1$ with transition probability $\mathbf{Pr}(\sigma_{\pi_n^{(s)}(t)}(t)=1)$ in \cref{E[r_k^n]}: No observation occurs, so the state $\tilde{\bm{\mu}}^n(t+1)$ remains unchanged.
    \item $\sigma_{\pi_n^{(s)}(t)}(t)=0$ and $r_{\pi_n^{(s)}(t)}^n(t)=1$ with transition probability $\mathbf{Pr}(\sigma_{\pi_n^{(s)}(t)}(t)=0)\tilde{\mu}_{\pi_n^{(s)}(t)}^n(t)$: $\tilde{\mu}_{\pi_n^{(s)}(t)}^n(t+1)$ of arm $k$ increases via the update rule in \cref{tilde_mu}, while $\tilde{\mu}^n_k(t+1)$ for all other arms remain unchanged.
    \item $\sigma_{\pi_n^{(s)}(t)}(t)=0$ and $r_k^n(t)=0$ with transition probability $\mathbf{Pr}(\sigma_{\pi_n^{(s)}(t)}(t)=0)(1-\tilde{\mu}_{\pi_n^{(s)}(t)}^n(t))$: $\tilde{\mu}_{\pi_n^{(s)}(t)}^n(t+1)$ of arm $k$ decreases via the update rule in \cref{tilde_mu}, while $\tilde{\mu}^n_k(t+1)$ of all other arms remain unchanged.
\end{itemize}

For ease of exposition, in this paper we focus on the pure strategies instead of mixed strategies of selfish players, where each player deterministically selects one arm at any given time (\!\!\cite{rosenski2016multi,boursier2020selfish}), as they are simpler to compute and implement. 
Our main results, including the PoA analysis in \Cref{thm:PoA_selfish} and mechanism design in \Cref{section5}, are also applicable to mixed strategies, in which players choose an arm probabilistically according to a predefined probability distribution. 
This is because pure and mixed strategies eventually converge to the same equilibrium outcome, resulting in the same number of players on each arm after convergence.

By observing \cref{pi_n_s(t)}, we find that selfish players only care about their own long-term rewards, overlooking the possible collisions to reduce rewards for other players on the same arm. Although the state space of $V_n^{(s)}(\tilde{\bm{\mu}}^n(t))$ increases exponentially with time gap $i$ in \cref{pi_n_s(t)}, we still derive the structural characterization of $\pi_n^{(s)}(t)$ later in \Cref{section4}. 

\begin{algorithm}[t]
\caption{Decision-making for the social planner under the socially optimal policy}
\small
\label{algo:social_optimum}
\begin{algorithmic}[1]
\STATE \textbf{Input:} $\mathbb{T},\mathbb{N},\mathbb{K}$;\
\STATE Initialize $\Tilde{\mu}_k(0)=\theta_k,c_k^n(0)=0$, $\forall k\in\mathbb{K}$ and $n\in\mathbb{N}$;\
\FOR{$t\in\mathbb{T}$}
\STATE Solve \cref{pi^*(t)} to obtain $\bm{\pi}^*(t)$;\
\STATE Let each player $n$ choose arm $\pi^*_n(t)$ and observe $r_k(t)$;\ 
\STATE Update $c_k^n(t+1)=c_k^n(t)+1$ for any $n\in\mathbb{N}$;\
\STATE Update $\Tilde{\mu}_k(t+1)$ by \cref{tilde_mu_opt};\
\ENDFOR
\end{algorithmic}
\end{algorithm}

\textbf{Socially optimal policy.} To compare selfish policy \cref{pi_n_s(t)} with the performance upper bound of strategy profile, we then formulate the optimization problem for the socially optimal policy.
In this scenario, all players listen to a centralized social planner, who follows the socially optimal policy for maximizing the long-term total reward for all players.
In addition, the social planner has access to all arm information observed by players, including $c_k^n(t)$ in \cref{c_k^n} and $r_k^n(t)$ in \cref{r_k^n} for all arms. 
Therefore, the social planner can estimate the global empirical mean reward $\tilde{\mu}_k(t+1)$ of arm $k$ as:
\begin{align}
    \tilde{\mu}_k(t+1)=\frac{\sum_{\tau=1}^t\sum_{n=1}^N\mathds{1}\{\pi_n(\tau)=k\}r^n_k(\tau)}{\sum_{n=1}^Nc_k^n(t+1)},\label{tilde_mu_opt}
\end{align}
Let $\tilde{\bm{\mu}}(t)$ denote the set of empirical mean rewards, containing $\tilde{\mu}_k(t+1)$ for each arm $k\in\mathbb{K}$.

Let $\bm{\pi}^*(t)$ denote the decision vector of all players under the socially optimal policy. Define $V^*(\Tilde{\bm{\mu}}(t))$ to be the long-term expected total reward of all players under the socially optimal policy. Then we similarly formulate the optimal optimization problem for the social planner as:
\begin{align}\label{pi^*(t)}
    V^*(\tilde{\bm{\mu}}(t))= \max_{\bm{\pi}^*(t)\in\mathbb{K}}& \Big\{\sum_{n=1}^N \tilde{r}^n_{\pi_n^*(t)}(t)\\&+\rho\mathbb{E}\Big[V^*(\tilde{\bm{\mu}}(t+1))\Big|r_{\pi_n^*(t)}^n(t),\forall n\Big]\Big\}.\notag
\end{align}
Unlike the distributed decision-making under the selfish policy \cref{pi_n_s(t)}, the socially optimal policy \cref{pi^*(t)} decides the policy set $\bm{\pi}^*(t)$ in a centralized way to maximize the sum of all players' long-term discounted rewards. It not only promptly explores different arms but also avoids collisions among players. 

Note that $\bm{\pi}^*(t)$ in \cref{pi^*(t)} is not unique, as any two players can exchange their arm decisions $\pi^*_n(t)$ without changing the total reward. 
With this useful property, we can flexibly recommend players to choose their most preferred arms from $\bm{\pi}^*(t)$ in our new CISP mechanism design, as discussed later in \Cref{section5}. 

\section{ANALYSIS OF POLICIES AND POA FOR CMAB GAMES}\label{section4}
In this section, we first provide threshold-based structures and convergence analysis for both selfish policy and socially optimal policy to understand and compare. Subsequently, we demonstrate that, as compared to the socially optimal policy in \cref{pi^*(t)}, players' selfish policy in \cref{pi_n_s(t)} results in an infinite $\text{PoA}$. 

In the following, for some positive constant $c_1$ and $c_2$, we define $x=\Omega(y)$ if $x>c_2|y|$ and $x=\mathcal{O}(y)$ if $x<c_1|y|$. We also denote $x=o(y)$ is $x/y\rightarrow 0$.



\subsection{Analysis of Selfish Policy}

In the following proposition, we derive the threshold-based structure that determines when each player switches arm decisions under the selfish policy \cref{pi_n_s(t)}.

\begin{proposition}\label{prop:selfish_threshold}
Under selfish policy \cref{pi_n_s(t)}, consider player $n\in\mathbb{N}_k(t-1)$, who chose arm $\pi_n^{(s)}(t-1)=k$ in the previous time slot $t-1$.
At time $t$, there exists a unique exploration threshold that determines whether player $n$ should switch to another arm $j\neq k$. This threshold depends on the following case-by-case scenarios: 
\begin{align}
    \mathcal{T}_{j,k}^n(t)\in \begin{cases}
        \Big[\tilde{r}_k^n(t)-\Delta \mu_{j,k}(t), \tilde{r}_k^n(t) \Big], \text{ if } c_j^n(t)<c_k^n(t),\\
        \Big[\tilde{r}_k^n(t), \tilde{r}_k^n(t)-\Delta \mu_{j,k}(t) \Big], \text{ if } c_j^n(t)\geq c_k^n(t),
    \end{cases}\label{threshold_T}
\end{align}
where $c_k^n(t)$ and $c_j^n(t)$ are the up-to-now exploration counts for arms $k$ and $j$, respectively, as given in \cref{c_k^n}, and $\tilde{r}_k^n(t)$ is the expected reward given in \cref{E[r_k^n]}, and
\begin{align}
\textstyle
    \Delta \mu_{j,k}(t)=\frac{(c_k^n(t)-c_j^n(t))(1-\tilde{r}_k^n(t))}{(\mathbb{E}\big[|\mathbb{N}_j(t-1)|\big|\tilde{\bm{\mu}}^n(t-1)\big]+1)(c_k^n(t)c_j^n(t)+c_j^n(t))}.\label{delta_mu_selfish}
\end{align}
Player $n$ will switch to arm $j$ if the expected reward of arm $j$ satisfies $\tilde{r}_j^n(t)>\mathcal{T}^n_{j,k}(t)$.
In this case, choosing arm $j$ strictly dominates staying with arm $k$ under selfish policy \cref{pi_n_s(t)}. 
The threshold $\mathcal{T}_{j,k}^n(t)$ in \cref{threshold_T} increases with $c_j^n(t)$. If $c_j^n(t)<c_k^n(t)$, $\mathcal{T}_{j,k}^n(t)$ decreases with discount factor $\rho$. Otherwise, $\mathcal{T}_{j,k}^n(t)$ increases with $\rho$.
\end{proposition}

The proof of \Cref{prop:selfish_threshold} is given in Appendix~B. \Cref{prop:selfish_threshold} provides a structural characterization of how non-myopic players behave dynamically under selfish policy \cref{pi_n_s(t)} in CMAB games. Note that such explicit policy analysis is not provided in the literature \cite{boursier2019sic,branzei2021multiplayer,sentenac2021decentralized,huang2023near,xu2023competing}.

If discount factor $\rho=0$, players become myopic and simply compare the one-shot expected rewards of arms $k$ and $j$ to make their current arm decisions in \cref{threshold_T}. In this case, we have $\mathcal{T}_{j,k}^n(t)=\tilde{r}_k^n(t)$.
While if $\rho>0$, players are non-myopic and focus on their own long-term rewards.
According to \cref{threshold_T}, there are two cases for player $n$ to decide its exploration threshold $\mathcal{T}^n_{j,k}(t)$, depending on its number of explorations of arm $k$ and arm $j$:
\begin{itemize}
    \item If player~$n$ has explored arm $k$ more than arm~$j$ (i.e., $c_j^n(t)<c_k^n(t)$), it is more inclined to switch from arm~$k$ to arm $j$, even if arm $j$ offers a lower immediate expected reward compared to the reward of sticking with arm $k$, i.e., $\tilde{r}_j^n(t)<\tilde{r}_k^n(t)$. 
    This preference arises because arm $j$, with less exploration than arm $k$, offers a positive exploration benefit for player $n$'s long-term reward, which is at most $\Delta \mu_{j,k}(t)$ in \cref{delta_mu_selfish}.
    \item If player~$n$ has explored arm $k$ less than arm~$j$ (i.e., $c_j^n(t)>c_k^n(t)$), it becomes less inclined to explore arm $j$ due to the negative exploration benefit ($\Delta \mu_{j,k}(t)<0$) compared to sticking with arm $k$ for its long-term reward. 
\end{itemize}

The maximum exploration benefit $\Delta \mu_{j,k}(t)$ in \cref{delta_mu_selfish} is achieved at $\rho=1$. Note that if $c_j^n(t)=0$ and $c_k^n(t)\neq 0$, the maximum exploration benefit $\Delta \mu_{j,k}(t)$ is arbitrarily large, implying that player $n$ will switch to arm $j$ without any exploration after exploring arm $n$ under $\rho=1$. 
As the exploration counts $c_k^n(t)$ and $c_j^n(t)$ for arms $k$ and $j$ increase, the maximum exploration benefit $\Delta \mu_{j,k}(t)$ in \cref{delta_mu_selfish} approaches zero, leading to $\mathcal{T}_{j,k}^n(t)=\tilde{r}_k^n(t)$. This indicates that player $n$ becomes myopic, focusing on maximizing its one-shot reward. 



As the exploration counts $c_j^n(t)$ of all arms increases, the exploration benefit in \cref{delta_mu_selfish} decreases to zero, and players converge to the Nash equilibrium (NE), denoted by $\Bar{\bm{\pi}}^{(s)}$. In $\Bar{\bm{\pi}}^{(s)}$, no player can unilaterally switch arms to improve its own reward. 
Let $\Bar{\bm{\pi}}^{(s)}_{-n}$ denote the arm decisions of the other $N-1$ players except player $n$.
Then we generally define the $\epsilon$-Nash equilibrium ($\epsilon$-NE, \cite{roughgarden2010algorithmic}) for our CMAB games.
\begin{definition}[$\epsilon$-Nash Equilibrium ($\epsilon$-NE)]\label{def:nash_equilibrium}
In CMAB games, an arm decision $\Bar{\pi}_n^{(s)}=k\in \mathbb{K}$ is an NE for player $n$ if for all $k'\in\mathbb{K}$ with $k'\neq k$,
\begin{align}\label{def_NE}
    \mathbb{E}\big[r_k^n(t)\big|\bar{\bm{\pi}}^{(s)}\big]\geq \mathbb{E}\big[r_{k'}^n(t)\big|{\pi}_n^{(s)}=k',\bar{\bm{\pi}}_{-n}^{(s)}\big]-\epsilon,
\end{align}
where $\epsilon=o(1)$.
\end{definition}
Note that when $\epsilon=0$, $\Bar{\pi}_n^{(s)}$ becomes the NE in \cref{def_NE}.
Based on \Cref{def:nash_equilibrium}, we define the convergence of selfish players in CMAB games.
\begin{definition}[Convergence]\label{def:convergence}
In CMAB games, the convergence refers to all players' arm decisions converging to $\epsilon$-NE $\Bar{\pi}_n^{(s)}$.
\end{definition}

Let $T^{(s)}$ denote the convergence time for all players to converge to the $\epsilon$-NE. Recall that $\eta$ is the minimal gap between the mean rewards of any two arms, i.e., $|\mu_i-\mu_j|>\eta$ for any $i,j\in\mathbb{K}$. 
We then analyze the order of $T^{(s)}$ in the following proposition.

\begin{proposition}\label{lemma:exploration_period_selfish}
Under the selfish policy in \cref{pi_n_s(t)}, for $\eta,\delta>0$, with probability at least $1-\delta$, all players will converge to equilibrium $\Bar{\bm{\pi}}^{(s)}$ in $T^{(s)}= \Omega\big(\frac{K}{\eta^2}\ln({\frac{KN}{\delta}})\big)$ time slots.
\end{proposition}
The proof of \Cref{lemma:exploration_period_selfish} is given in Appendix~C. Under the selfish policy in \cref{pi_n_s(t)}, each player $n\in\mathbb{N}$ subsequently explores arms under \cref{threshold_T} until converging to $\Bar{\pi}_n^{(s)}$. The likelihood of collisions per time slot requires additional time slots for players to choose each arm. In the next subsection, we will compare $T^{(s)}$ with $T^*$ under the social optimum to show the exploration loss of the selfish policy.

\subsection{Analysis of Socially Optimal Policy}

Before analyzing the exploration threshold for the socially optimal policy, we first demonstrate that the socially optimal policy always avoids collisions among players in the next lemma.
\begin{lemma}\label{lemma:optimal_nocollision}
Under the socially optimal policy in \cref{pi^*(t)}, at any time $t\in\mathbb{T}$, the number of players choosing each arm $k\in\mathbb{K}$ satisfies $|\mathbb{N}_k(t)|\leq 1$.
\end{lemma}
The proof of \Cref{lemma:optimal_nocollision} is given in Appendix~D. 
The socially optimal policy ensures $|\mathbb{N}_k(t)|\leq 1$ to avoid collisions among players, thereby maximizing the total rewards for all players in \cref{pi^*(t)}.

Let $\mathcal{T}^*_{j,k}(t)$ denote the exploration threshold of arm $j$ under socially optimal policy \cref{pi^*(t)}. Unlike $\mathcal{T}^n_{j,k}(t)$ in \cref{threshold_T}, which explores arms in a distributed way, the socially optimal policy decides arms in a centralized way. 
Based on \Cref{lemma:optimal_nocollision}, if $\tilde{\mu}_j(t)<\mathcal{T}^*_{j,k}(t)$, we similarly characterize the solution of the socially optimal policy \cref{pi^*(t)} in the following proposition.
\begin{proposition}\label{prop:optimal_threshold}
Under socially optimal policy \cref{pi^*(t)}, there exists a common exploration threshold in the following case-by-case ranges for player $n$ to consider arm $j\neq k$ with $|\mathbb{N}_j(t-1)|=0$ at time $t$:  
\begin{align}
    \mathcal{T}^*_{j,k}(t)\in \begin{cases}
        \big[\tilde{\mu}_k(t)-\Delta \mu^*_{j,k}(t), \tilde{\mu}_k(t)\big], \text{ if } c_j(t)<c_k(t),\\
        \big[\tilde{\mu}_k(t), \tilde{\mu}_k(t)-\Delta \mu_{j,k}^*(t) \big], \text{ if } c_j(t)\geq c_k(t),
    \end{cases}\label{threshold_T*}
\end{align}
depending on up-to-now exploration numbers $c_k^n(t)$ and $c_j^n(t)$ between the two arms $k$ and $j$, where 
\begin{align}
    \Delta \mu_{j,k}^*(t)=\frac{(c_k(t)-c_j(t))(1-\tilde{\mu}_k(t))}{c_k(t)c_j(t)+c_j(t)}. \label{delta_mu_optimal}
\end{align}
That is, if the empirical mean reward of arm $j$ satisfies $\tilde{\mu}_j(t)<\mathcal{T}_{j,k}(t)$, then $\pi_n^*(t)=k$ strictly dominates $\pi_n^*(t)=j$ for player $n$. 
The exploration threshold $\mathcal{T}^*_{j,k}(t)$ in \cref{threshold_T*} increases with $c_j(t)$. Additionally, if $c_j(t)<c_k(t)$, $\mathcal{T}^*_{j,k}(t)$ decreases with discount factor $\rho$. Otherwise, $\mathcal{T}^*_{j,k}(t)$ increases with $\rho$.
\end{proposition}
The proof of \Cref{prop:optimal_threshold} is given in Appendix~E.
The optimal exploration threshold $\mathcal{T}^*_{j,k}(t)$ in \cref{threshold_T*} is similarly derived as $\mathcal{T}_{j,k}^n(t)$, by letting $|\mathbb{N}_k(t-1)|=1$ and $\mathbb{E}[|\mathbb{N}_j(t-1)|\big|\tilde{\bm{\mu}}^n(t-1)]=0$ in \cref{threshold_T} to avoid collision, based on \Cref{lemma:optimal_nocollision}.
According to \cref{threshold_T*}, to avoid collisions among players, the socially optimal policy only compares arm $k$ to another arm~$j$ without player exploration at last time $t-1$, i.e., $|\mathbb{N}_j(t-1)|=0$. This is different from the selfish threshold $\mathcal{T}_{j,k}^n(t)$ in \cref{threshold_T} that compares to any arm $j\neq k$.

According to \cref{threshold_T*}, as $c_k(t)$ and $c_j(t)$ increase to make the maximum exploration benefit $\Delta \mu^*_{j,k}(t)$ in \cref{delta_mu_selfish} to approach zero, the exploration threshold $\mathcal{T}^*_{j,k}(t)$ becomes myopic $\tilde{\mu}_k(t)$. 
Meanwhile, the empirical mean reward $\tilde{\mu}_k(t)$ converges for each arm~$k$.

Let $\bar{\bm{\pi}}^*$ denote the steady decision set after convergence under the socially optimal policy \cref{pi^*(t)}. 
Based on \Cref{lemma:optimal_nocollision} and \Cref{prop:optimal_threshold}, in the steady decision $\Bar{\bm{\pi}}^*$, the socially optimal policy always chooses the top $N$ arms to maximize the long-term total reward for $N$ players. 
In the next proposition, we similarly analyze the convergence time under the socially optimal policy \cref{pi^*(t)}, denoted by $T^*$.
\begin{proposition}\label{lemma:exploration_stage_optimal}
Under socially optimal policy in \cref{pi^*(t)}, for $\eta,\delta>0$, with probability at least $1-\delta$, all players will converge to the optimal steady decision $\Bar{\bm{\pi}}^*$ in $T^*= \mathcal{O}\big(\frac{K}{N\eta^2}\ln{(\frac{K}{\delta})}\big)$ time slots.
\end{proposition}
The proof of \Cref{lemma:exploration_stage_optimal} is given in Appendix~F. Different from selfish policy's subsequent explorations with potential collisions in \Cref{lemma:exploration_period_selfish}, socially optimal policy \cref{pi^*(t)} recommends $N$ players simultaneously explore $N$ different arms to avoid collisions. Therefore, the convergence time is greatly reduced to $\mathcal{O}(\frac{K}{N\eta^2}\ln{(\frac{K}{\delta})})$, which is still an upper bound. 
Its comparison with \Cref{lemma:exploration_period_selfish} also tells the importance of designing incentive mechanisms among selfish players to accelerate the convergence process. 

In the next subsection, we analytically compare the long-term total rewards of the two policies to show the huge efficiency loss of the selfish policy.

\subsection{PoA Analysis}
Following \cite{koutsoupias1999worst}, we define the price of anarchy (PoA) to be the maximum ratio between the long-term discounted rewards under socially optimal policy in \cref{pi^*(t)} and selfish policy in \cref{pi_n_s(t)}:
\begin{align}
    \text{PoA}=\max_{\substack{\bm{\mu},\bm{\theta}^n\\ \rho, N,K}} \frac{\sum_{t=1}^\infty \rho^{t-1}\sum_{n=1}^N \tilde{r}^n_{\pi^*_n(t)}(t)}{\sum_{t=1}^\infty \rho^{t-1} \sum_{n=1}^N \tilde{r}^n_{\pi^{(s)}_n(t)}(t)},\label{PoA_definition}
\end{align}
which is always larger than $1$. Compared to the long-term total reward under the socially optimal policy, a higher PoA in \cref{PoA_definition}signifies greater efficiency loss under the selfish policy.


Next, we rigorously derive PoA caused by the selfish policy. Without loss of generality, to ease our following discussions, we sort the arms according to $\mu_1>\mu_2>\cdots>\mu_K$.
\begin{theorem}\label{thm:PoA_selfish}
As compared to socially optimal policy in \cref{pi^*(t)}, players' selfish policy in \cref{pi_n_s(t)} leads to 
\begin{align}
    \text{PoA}=1+\frac{\sum_{k=2}^{N} \mu_k}{\mu_1}, \label{PoA^s}
\end{align}
which is achieved when $\rho\rightarrow 0,\mu_1\rightarrow 1$ and $\theta_1^n>N\theta_k^n$ for any player~$n$ at any arm $k\in\{1,\cdots, N\}$. PoA in \cref{PoA^s} approaches infinity as $N\rightarrow \infty, K\rightarrow \infty$ and $\mu_k\rightarrow 1$ for any arm $k\in\{2,\cdots, N\}$.
\end{theorem}

The proof of \Cref{thm:PoA_selfish} is given in Appendix~G. Intuitively, the worst-case scenario happens when all selfish players stick with arm 1 ($\mu_1\rightarrow 1$ and $\theta_1^n>N\theta_k^n$) from the initial time $t=1$, leading to maximum collisions among all players and minimum exploration of other arms (under $\rho\rightarrow 0$). While the socially optimal policy lets each player choose one of the top $N$ arms to avoid collisions. In this case, as the actual mean reward $\mu_k\rightarrow 1$ for any arm $k\in\{2,\cdots, N\}$, the long-term total reward for all players is greatly increased. 
Note that only in the extreme case of only $N=1$ player, the selfish policy becomes the same as the social optimum with $\text{PoA}=1$ in \cref{PoA^s}.

According to Theorem $\ref{thm:PoA_selfish}$, PoA caused by the selfish policy can approach infinity, leading to arbitrary bad efficiency loss compared to the social optimum. Consequently, it is necessary to design an efficient mechanism to improve system performance and accelerate convergence for CMAB games.

\section{COMBINED INFORMATIONAL AND SIDE-PAYMENT (CISP) MECHANISM}\label{section5}
In this section, we aim to design an efficient mechanism to reduce the $\text{PoA}=\infty$ in \Cref{thm:PoA_selfish} to the optimum $1$. We first follow mechanism design literature \cite{borgers2015introduction} to formally define informational (non-monetary) mechanisms below.
\begin{definition}[Informational mechanisms \cite{borgers2015introduction}]\label{def:info_mechanism}
    An informational mechanism defines a Bayesian game in which the social planner decides how information is shared or withheld among selfish players to influence their decisions and maximize the expected total reward as defined in \cref{pi^*(t)}.
\end{definition}
Based on \Cref{def:info_mechanism}, we next prove that any informational mechanism in the literature (e.g., Bayesian persuasion \cite{kremer2014implementing,mansour2022bayesian,li2024optimize,li2025analyze}) cannot reduce PoA for selfish policy in our CMAB games. 
\begin{lemma}\label{lemma:information_mechanism}
Any informational mechanism cannot reduce $\text{PoA}=\infty$ in \Cref{thm:PoA_selfish} to a finite value for non-myopic selfish players.
\end{lemma}
The proof of \Cref{lemma:information_mechanism} is given in Appendix~H.
To prove this lemma, we consider the same worst case with maximum collisions as in \Cref{thm:PoA_selfish}. Given $\rho\rightarrow 0,\mu_1\rightarrow 1$ and $\theta_1^n>N\theta_k^n$ for any player~$n$ at any arm $k\in\{1,\cdots, N\}$, all selfish players will stick with the best arm 1 to exploit the largest long-term reward for themselves since the initial time $t=1$. 
In this case, any existing informational mechanisms, such as information hiding, partial disclosure, or private recommendations in \cite{mansour2022bayesian,tavafoghi2017informational,li2023congestion}, cannot influence their arm decisions, still resulting in $\text{PoA}=\infty$ as \Cref{thm:PoA_selfish}.

Furthermore, to design an informational mechanism, the social planner needs to collect arm information from players \cite{borgers2015introduction,mansour2020bayesian}, who as the information sources may strategically misreport their private observations to improve their own long-term benefits. The non-myopic nature of selfish players adds difficulty to the mechanism design. In contrast, studies such as \cite{mansour2022bayesian,tavafoghi2017informational,li2023congestion} only consider myopic players who are willing to share information truthfully in one shot.
As the social planner lacks information on each arm, existing monetary mechanisms to regulate myopic players cannot work, either (e.g., \cite{frazier2014incentivizing,che2018recommender}). This is because non-myopic players may misreport their private information to lower the charges or even earn rewards from the social planner.
Besides, players have different arm choices/observations in the past, which further adds difficulty to the monetary mechanism design.

Given \Cref{lemma:information_mechanism} and the necessity to have truthful information for guiding side-payment, we turn to jointly combine informational and side-payment incentives to design a new incentive-compatible mechanism for regulating CMAB games. We first follow the existing literature (e.g., \cite{myerson1979incentive}) to define incentive compatibility below.
\begin{definition}[Incentive compatibility \cite{myerson1979incentive}]\label{def:IC}
A mechanism is incentive compatible if, for each player, reporting its true reward observations maximizes its reward, assuming that other players do the same.
\end{definition}
The social planner may charge a fee from some players and reward some others over time, for altering their reward objectives in \cref{pi_n_s(t)}.
According to \cite{borgers2015introduction}, it is critical for a mechanism to ensure ex-post budget balance for the social planner's sustainable operation, which means the social planner’s budget should be no less than zero at any time in the time horizon. 
Furthermore, we want to ensure individual rationality for all players to participate in CMAB in the long run. Then we follow \cite{gode1993allocative} to formally define individual rationality below:
\begin{definition}[Individual rationality \cite{gode1993allocative}]\label{def:IR}
A player is individually rational if its reward from participating is at least as high as its utility from not participating (or from an outside option).
\end{definition}


\begin{algorithm}[t]
\caption{Combined informational and side-payment (CISP) mechanism}
\small
\label{algo:CISP}
\begin{algorithmic}[1]
\STATE \textbf{Input:} $\mathbb{N}, \mathbb{K}, \mathbb{T},\Psi\gg 1$;\
\STATE \textbf{Initialize} $\tilde{\bm{\mu}}(0), \mathbb{N}_k(0)$;\
\FOR{$t\in\mathbb{T}$}
\STATE \textbf{\# Step 1: Information aggregation}
\FOR{$n\in\{1,\cdots, N\}$}
\STATE Aggregate player $n$'s selfish policy $\pi_n^{(s)}(t)$, empirical mean reward set $\tilde{\bm{\mu}}^n(t)$, and exploration number $c_k^n(t)$ for each arm $k\in\mathbb{K}$;\ \label{line_aggregate}
\ENDFOR
\STATE Calculate $|\mathbb{N}_k(t)|$ for any $k$ and $\tilde{\bm{\mu}}(t)$ by \cref{tilde_mu};\
\STATE Solve \cref{pi^*(t)} to obtain $\mathbb{K}^*(t)$;\
\STATE \textbf{\# Step 2: Informational incentives}
\FOR{any player $n$ with $\pi_n^{(s)}(t)\notin \mathbb{K}^*(t)$}
\STATE Privately recommends player~$n$ to choose arm $\pi_n^*(t)=k$, where $k\in \mathbb{K}^*(t)$ with $|\mathbb{N}_k(t)|=0$ and $\Tilde{\mu}_k(t)>\mathcal{T}_{k,\pi_n^{(s)}(t)}^n(t)$ with $\mathcal{T}_{k,\pi_n^{(s)}(t)}^n(t)$ in \cref{threshold_T} under $\Tilde{\bm{\mu}}(t)$;\ \label{line_informational}
\ENDFOR
\STATE \textbf{\# Step 3: Side-payment incentives}
\FOR{any $i\in \mathbb{K}^*(t)$ with $|\mathbb{N}_i(t)|>1$}
\STATE Recommend $\pi_n^*(t)=i$ to player $n$, where $n=\arg \max_{h\in \mathbb{N}_i(t)}\big\{\tilde{\mu}_i^h(t)\big\}$, and charges it a payment
    \begin{align}
        p_i(t)=\frac{|\mathbb{N}_i(t)|-1}{|\mathbb{N}_i(t)|}\tilde{\mu}_i^n(t). \label{payment_i(t)}
    \end{align}
\ENDFOR
\FOR{any $j\in\mathbb{K}^*(t)$ with $|\mathbb{N}_j(t)|=0$}
\STATE Recommend $\pi_l^*(t)=j$ to player $l\in\mathbb{N}^{-n}$, where
\begin{align}
        l=\arg \min_{h\in \mathbb{N}^{-n}}\left\{\frac{\tilde{\mu}_{\pi^{(s)}_h(t)}^h(t)}{|\mathbb{N}_{\pi^{(s)}_h(t)}(t)|}-\tilde{\mu}_j^h(t)\right\}\label{reward_n}
    \end{align}
    with $|\mathbb{N}_{\pi^{(s)}_l(t)}|>1$ and $\mathbb{N}^{-n}$ to be the set of rest players not recommended in steps 2 or~3;\
\STATE Provides a reward
    \begin{align}
     p_j(t)=\frac{\tilde{\mu}_{\pi^{(s)}_l(t)}^n(t)}{|\mathbb{N}_{\pi^{(s)}_l(t)}(t)|}-\tilde{\mu}_j^l(t)\label{reward_j}
    \end{align} 
    to player $l$ who follows the optimal recommendation $\pi_l^*(t)=j$, where $n=\arg \max_{h\in \mathbb{N}_{\pi_l^{(s)}(t)}(t)}\big\{\tilde{\mu}_{\pi_l^{(s)}(t)}^h(t)\big\}$;\
\STATE Announce payment $p_{\pi_l^{(s)}(t)}(t)$ derived in \cref{payment_i(t)} for charging player $l$ if it sticks with its selfish decision $\pi_l^{(s)}(t)$;\ \label{line_announce}
\ENDFOR
\STATE \textbf{\# Step 4: Verification with penalty}
\STATE Observe each player $n$'s final arm decision $\pi^{(\$)}_n(t)$;\ 
\FOR{any player $n$ with $\pi^{(\$)}_n(t)\neq \pi^*_n(t)$}
\STATE Charges a penalty $\Psi$ from player $n$;\ \label{line_charge}
\STATE Use $\Psi$ to compensate any player $l$ with $\pi^{(\$)}_n(t)=\pi^*_l(t)$;\ \label{line_compensate}
\ENDFOR
\ENDFOR
\end{algorithmic}
\end{algorithm}

Based on \Cref{def:IC} and \Cref{def:IR}, we are ready to formally propose our combined informational and side-payment (CISP) mechanism in \Cref{algo:CISP}. 
For our CISP mechanism, we define an arm set $\mathbb{K}^*(t)$ to contain the $N$ arms chosen by the socially optimal policy in \cref{pi^*(t)}, based on \Cref{lemma:optimal_nocollision}. Consistent with the existing MAB literature (e.g., \cite{wu2019learning,macault2022social,li2023congestion}), we assume that the social planner can observe all players' arm decisions after pulling their chosen arms. For example, in spectrum sharing applications, a social planner can monitor players' received signals across all channels to detect spectrum usage and manage access effectively ([39], [40]). Additionally, in transportation networks \cite{li2023congestion}, navigation platforms can track users' path (arm) decisions through GPS data. Even in scenarios where direct observation of players' decisions is unavailable, the social planner can leverage reports from other players—particularly those involved in collisions—to identify deviations and impose corresponding penalties.

According to \Cref{algo:CISP}, our CISP mechanism first asks players to report their latest private information and then summarizes reports in an aggregated probability set $\Tilde{\bm{\mu}}(t)$ (Line~\ref{line_aggregate}). This creates information asymmetry between the social planner and an individual player, and our CISP can use $\Tilde{\bm{\mu}}(t)$ to derive the optimal arm set $\mathbb{K}^*(t)$ and design incentives. Note that the high penalty $\Psi\gg 1$ for penalizing players' deviation from optimal policy $\bm{\pi}^*(t)$ in step 4 ensures players' truthful reporting in the first step. In step 2, for any player $n$ preferring $\pi_n^{(s)}(t)\notin \mathbb{K}^*(t)$, our CISP recommends it to change its decision to another arm $k\in \mathbb{K}^*(t)$ (Line~\ref{line_informational}). As the two exploration threshold bounds in \cref{threshold_T} and \cref{threshold_T*} satisfy $\mathcal{T}^n_{k,\pi_n^{(s)}(t)}(t)\leq \mathcal{T}^*_{k,\pi_n^{(s)}(t)}(t)$ under the same empirical mean reward $\Tilde{\bm{\mu}}(t)$, this optimal recommendation $\pi_n^*(t)$ leads to a higher long-term reward for player $n$, and it will follow it and truthfully report its observations there.

After step 2, all players' arm decisions belong to the optimal arm set $\mathbb{K}^*(t)$, and any informational incentive alone can no longer change their arm competition. 
Therefore, the social planner needs to design monetary incentives to make $|\mathbb{N}_k(t)|=1$ for any arm $k\in\mathbb{K}^*$. 
To realize this goal, in step 3, our CISP mechanism charges the payment $p_i(t)$ in \cref{payment_i(t)} from player $n\in\mathbb{N}_i(t)$ with the highest $\Tilde{\mu}_i^n(t)$, without changing its selfish arm decision~$i$. With the balance, for the rest players in $\mathbb{N}^{-n}$ without any recommendation yet, CISP rewards feasible $p_j(t)$ in \cref{reward_j} to persuade player $l\in \mathbb{N}^{-n}$ in \cref{reward_n} to
another arm $j\in\mathbb{K}^*(t)$ with $|\mathbb{N}_j(t)|=0$. Besides, our CISP mechanism also announces the payment $p_{\pi_l^{(s)}(t)}(t)$ derived in \cref{payment_i(t)} for tolling arm $\pi_l^{(s)}(t)$ in case they stick with their selfish arm decisions~$\pi_l^{(s)}(t)$ (Line~\ref{line_announce}).

Under the incentives in steps 2 and 3, players will make their final arm decisions, denoted as $\bm{\pi}^{(\$)}(t)$. Then in step~4, our CISP mechanism examines whether each player $n$'s final arm decision $\pi_n^{(\$)}(t)$ is the same as its recommended $\pi^*_n(t)$ or not. If player $n$ deviates, it means it misreported its private information in step~1, and it will face a huge penalty $\Psi\gg 1$ (Line~\ref{line_charge}). Afterwards, the social planner will use this budget $\Psi$ to compensate player $l$ who followed the optimal recommendation $\pi^*_l(t)=\pi^{(\$)}_n(t)$ in steps 2 or~3 but finally experienced a collision due to player $n$'s deviation (Line~\ref{line_compensate}). 
Under such a potential penalty, all players would not misreport their private information in step 1, and thus always follow optimal recommendation $\pi_n^*(t)$ under our CISP mechanism.

In the next proposition, besides players' incentive compatibility, we also prove their long-run participation.  
\begin{proposition}\label{prop:IC_IR}
Our CISP mechanism in \Cref{algo:CISP} ensures incentive compatibility and individual rationality for all players.
\end{proposition} 
The proof of \Cref{prop:IC_IR} is given in Appendix~I.
Then in the next lemma, we prove the social planner's ex-post budget balance under our CISP mechanism.

\begin{lemma}\label{lemma:BB}
Our CISP mechanism in \Cref{def:info_mechanism} keeps ex-post budget balanced for the social planner at any time.
\end{lemma}
The proof of \Cref{lemma:BB} is given in Appendix~K.
Thanks to our CISP, the ex-post budget balance for the social planner ensures its sustainable operations, and players are willing to participate and truthfully report in the long run.
Finally, we are ready to prove that our CISP achieves the optimal PoA and convergence order in the next theorem.
\begin{theorem}\label{thm:RSP_PoA}
Our CISP mechanism in \Cref{algo:CISP} successfully reduces the infinite PoA in \cref{PoA^s} caused by the selfish policy to $\text{PoA}=1$ at the optimum. The convergence time under the CISP mechanism matches that of the socially optimal policy, which is given by $\mathcal{O}\big(\frac{K}{N\eta^2}\ln{(\frac{K}{\delta})}\big)$.
\end{theorem}

The proof of \Cref{thm:RSP_PoA} is given in Appendix~J. Based on \Cref{prop:IC_IR}, players are incentive-compatible to always report truthfully and follow the optimal recommendations under our CISP mechanism. Therefore, our CISP mechanism successfully changes the sub-optimal player numbers $|\mathbb{N}_i(t)|>1$ and $|\mathbb{N}_j(t)|=0$ to $|\mathbb{N}_i(t)|=1$ and $|\mathbb{N}_j(t)|=1$ for any arm $i,j \in \mathbb{K}^*(t)$ all the time, leading to the optimum $\text{PoA}=1$ and convergence time $\mathcal{O}\big(\frac{K}{N\eta^2}\ln{(\frac{K}{\delta})}\big)$.

\section{EXPERIMENTS VERIFICATION}\label{section_simulation}
 
In addition to theoretical analysis in Sections~\ref{section4} and \ref{section5}, we further conduct simulation experiments to check the average-case performances of the selfish policy and our CISP mechanism as compared to the socially optimal policy.

\begin{figure}[t]
    \centering
    \includegraphics[width=0.43\textwidth]{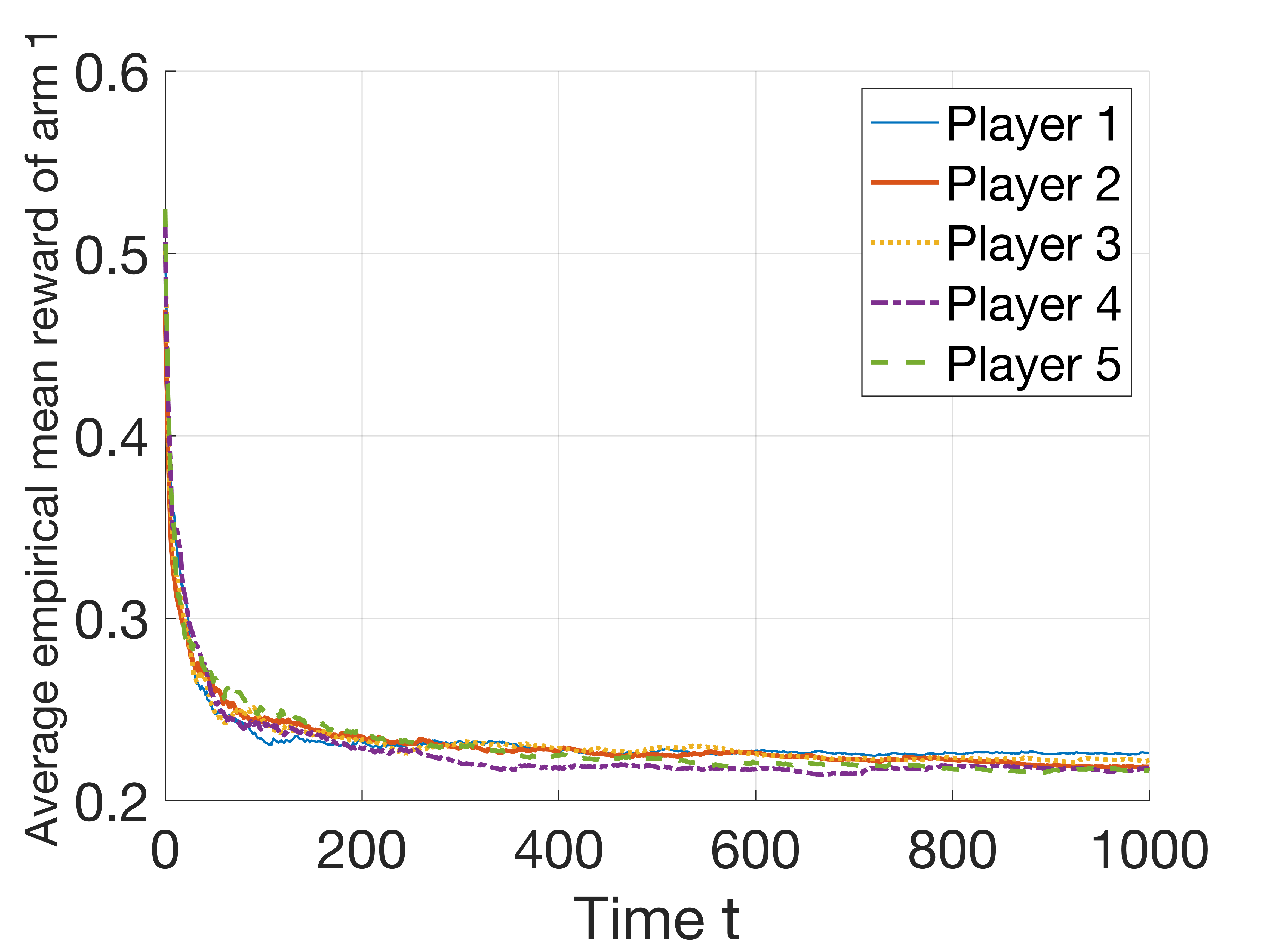}
    \caption{Comparison of $N=5$ players' average empirical mean rewards of arm 1 under selfish policy \cref{pi_n_s(t)}.}
    \label{fig:avg_reward}
\end{figure}

In the first experiment, we examine the dynamics of each player's average empirical mean reward for one of the $K$ arms to verify the assumption that each player reasonably believes that other players symmetrically have the same expected empirical mean reward $\tilde{\bm{\mu}}^n(t)$ from $t=0$. 
We consider $N=5$ transmitters (players) sharing $K=8$ resources (arms) to transmit data in a resource-sharing system over a finite time horizon of $T=1000$. 
We set $\rho=0.95,\eta=0.01,\delta=0.1$, with the probability of good conditions (mean reward set) given by $\bm{\mu}=[0.22\ 0.12\ 0.98\ 0.11\ 0.09\ 0.08\ 0.14\ 0.11]$. 
We randomly initialize $\theta_k^n$ from interval $[0,1]$ for any $k$ and $n$. 
The experiment is repeated $100$ times, and we compute the average of arm 1's empirical mean reward for each player $n\in[N]$. 
As depicted in Fig.~\ref{fig:avg_reward}, even though players' initial beliefs of arm 1 differ, their average dynamics are very close and will eventually approach its actual value $\mu_1=0.22$. This is consistent with Lemma~\ref{lemma:equal_expect}.

In the second experiment, we evaluate the learning efficiency of the selfish policy, the socially optimal policy, and our CISP mechanism. As a benchmark, we also examine the information hiding mechanism that is commonly used in the existing literature on Bayesian mechanism design (e.g., \cite{tavafoghi2017informational,li2023congestion}).
For the selfish policy in \cref{pi_n_s(t)}, we measure the average learning error of arm mean rewards for all players as follows:
\begin{align}
    \varepsilon^{(s)}(t)=\frac{1}{N\cdot K}\sum_{n=1}^N\|\bm{\mu}-\tilde{\bm{\mu}}^n(t)\|_2,
\end{align}
where $\|\bm{\mu}-\tilde{\bm{\mu}}^n(t)\|_2$ represents the $\ell$-2 norm of the difference between the player $n$'s actual mean reward $\bm{\mu}$ and the empirical mean reward set $\tilde{\bm{\mu}}^n(t)$ at time $t$. 
Similarly, we define $\varepsilon^{\emptyset}(t),\varepsilon^{(\text{CISP})}(t)$, and $\varepsilon^*(t)$ as the average learning errors of the information-hiding mechanism (e.g., \cite{tavafoghi2017informational,li2023congestion}), our CISP mechanism in \Cref{def:info_mechanism}, and the socially optimal policy in \cref{pi^*(t)}, respectively.

\begin{figure}[t]
    \centering
    \includegraphics[width=0.43\textwidth]{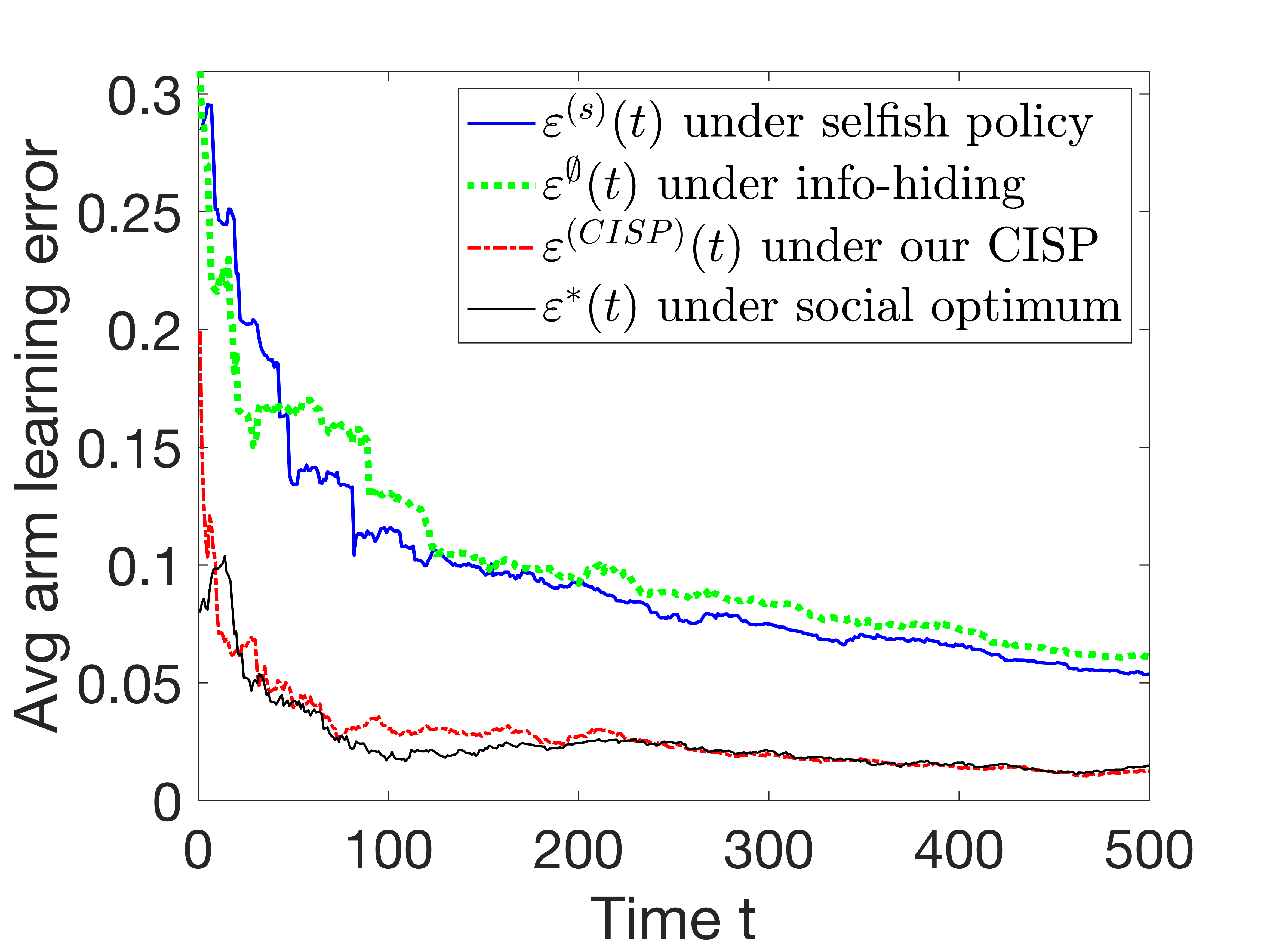}
    \caption{Comparison of average learning errors, under selfish policy \cref{pi_n_s(t)}, information-hiding mechanism (\!\!\cite{tavafoghi2017informational,li2023congestion}), our CISP mechanism in \Cref{def:info_mechanism}, and the socially optimal policy \cref{pi^*(t)}.}
    \label{fig:learning_error}
\end{figure}

We consider $N=8$ transmitters sharing $K=12$ resources to transmit data in a resource-sharing system over a finite time horizon of $T=500$. We set $\rho=0.95,\eta=0.01,\delta=0.1$, with the probability of good conditions (mean reward set) give by $\bm{\mu}=[0.22\ 0.12\ 0.98\ 0.11\ 0.09\ 0.08\ 0.14\ 0.11\ 0.09\ 0.08\ 0.14\ 0.11]$.
We initialize $\theta_k^n=0.5$ for any $k$ and $n$. In Fig.~\ref{fig:learning_error}, we plot the dynamics of the four learning errors $\varepsilon^{(s)}(t),\varepsilon^{\emptyset}(t),\varepsilon^{(\text{CISP})}(t)$, and $\varepsilon^*(t)$, over time $t\in\{1,\cdots, 500\}$. 
This figure shows that both $\varepsilon^{(s)}(t)$ under the selfish policy and $\varepsilon^{\emptyset}(t)$ under the hiding mechanism remain above $0.05$ after $t=500$ iterations and do not converge. Like the selfish policy, the information-hiding mechanism fails to regulate players' inefficient arm exploration under competition and collision. 
In contrast, our CISP mechanism effectively reduces the error, achieving $\varepsilon^{(\text{CISP})}(t)<0.02$ by $t=250$, closely approximating $\varepsilon^*(t)$ under the socially optimal policy. This outcome aligns with our \Cref{thm:RSP_PoA}.

\begin{figure}[t]
    \centering
    \subfigure[Discount factor $\rho=0.05$.]{\label{subfig:avg_rho0.05}\includegraphics[width=0.43\textwidth]{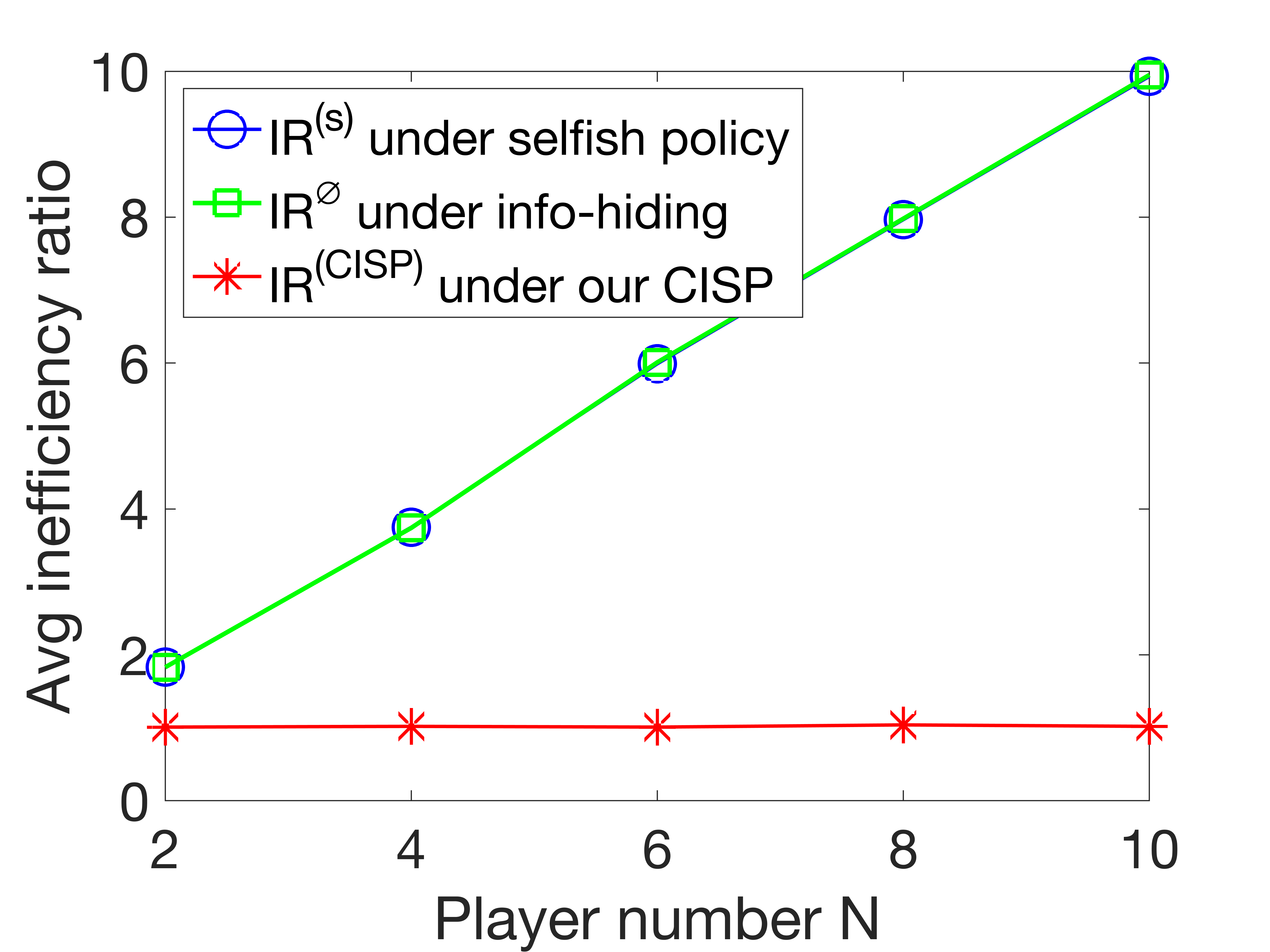}} 
    \hspace{0.2cm}
    \subfigure[Discount factor $\rho=0.95$]{\label{subfig:avg_rho0.95}\includegraphics[width=0.43\textwidth]{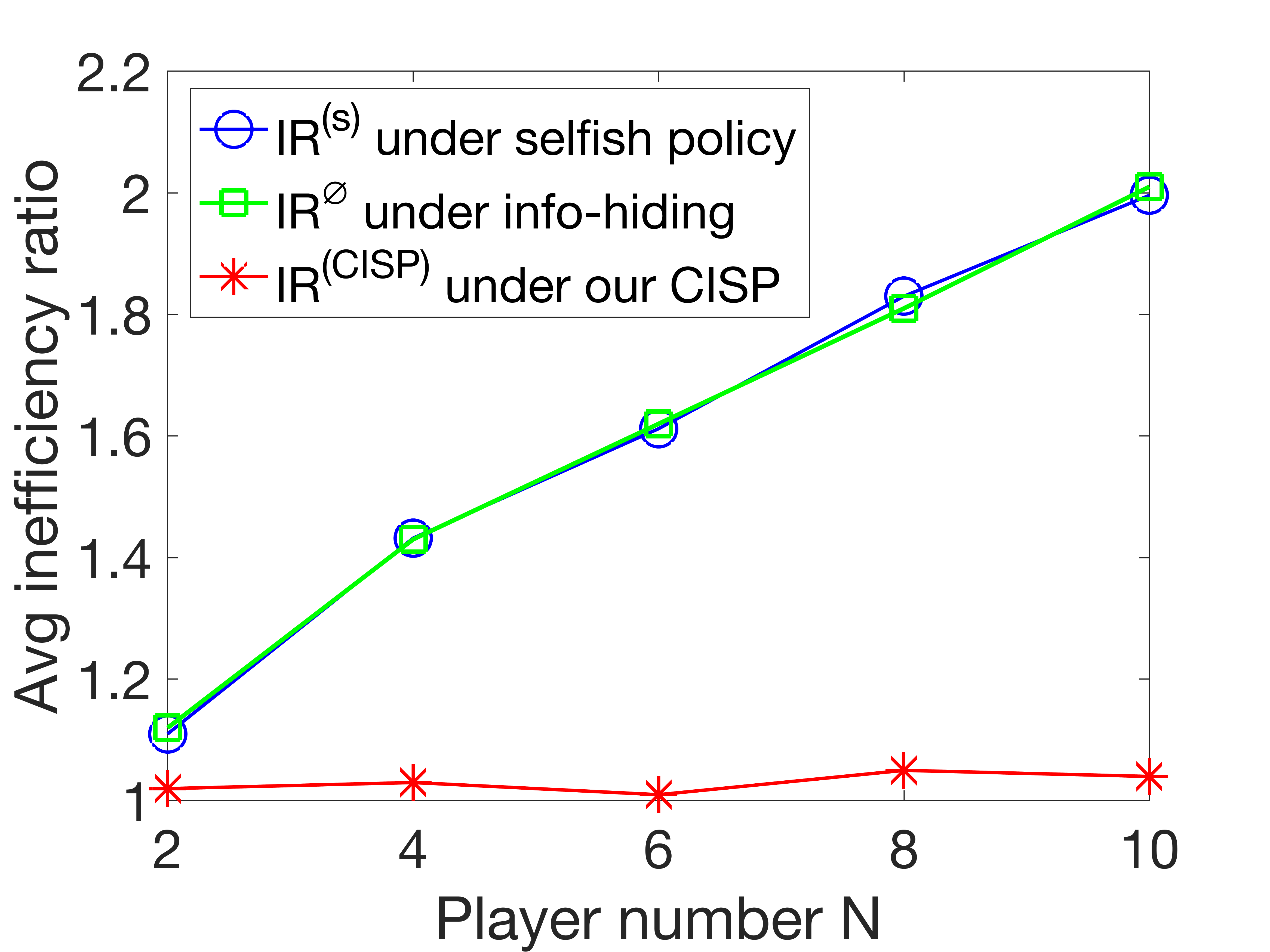}} 
    \caption{Comparison of average inefficiency ratios caused by selfish policy, information-hiding, and our CISP mechanism. We vary the number of players $N\in\{2,4,6,8,10\}$.}
    \label{fig:avg_ineff}
\end{figure}

Next, we examine the long-term social rewards of the selfish policy, information-hiding mechanism, and our CISP mechanism, compared to the socially optimal policy. Similar to the PoA ratio definition for the worst-case in \cref{PoA_definition}, we defined the average inefficiency ratios for long-term social reward caused by the selfish policy as
\begin{align*}
    \text{IR}^{(s)}= \frac{\sum_{t=1}^T \rho^{t-1}\sum_{n=1}^N \tilde{r}^n_{\pi^*_n(t)}(t)}{\sum_{t=1}^T \rho^{t-1} \sum_{n=1}^N \tilde{r}^n_{\pi^{(s)}_n(t)}(t)},
\end{align*}
where $T$ is the finite time horizon. Let $\text{IR}^{\emptyset}$ and $\text{IR}^{(\text{CISP})}$ similarly denote the average inefficiency ratios caused by the hiding mechanism and our CISP mechanism, respectively. 

In the third experiment in Fig.~\ref{fig:avg_ineff}, we set $K=12,T=2000, \eta=0.01,\delta=0.1$, and vary $N\in\{2,4,6,8,10\}$. Initially, for each player $n\in\mathbb{N}$, we fairly set $\theta_1^n=0.99$ and $\theta_k^n=0.05$ for any $k\neq 1$. We also examine two sub-experiments with different discount factors, $\rho_1=0.05$ and $\rho_2=0.95$. The actual mean rewards for these sub-experiments are set as:
\begin{align*}
    \bm{\mu}_1=[&0.99\ 0.95\ 0.94\ 0.97\ 0.98\ 0.98\ \\ &0.94\ 0.93\ 0.92\ 0.94\ 0.95\ 0.94],\\
    \bm{\mu}_2=[&0.99\ 0.24\ 0.24\ 0.24\ 0.24\ 0.08\ \\&0.24\ 0.23\ 0.22\ 0.24\ 0.15\ 0.24].
\end{align*}
We generally set $\rho_1=0.05$ and $\theta_1^n=0.99$ to assess the near worst-case scenario as described in \Cref{thm:PoA_selfish}, where no exploration of any arm $k\neq 1$ occurs (resulting in maximum performance loss as $\rho\rightarrow 0$). We set $\rho_2=0.95$ to evaluate more typical average-case performance.

Under these settings, we conducted $50$ experiments to average the inefficiency ratios to plot Fig.~\ref{fig:avg_ineff}.
In Fig.~\ref{subfig:avg_rho0.05}, we observe that $\text{IR}^{(s)}$ under the selfish policy increases with $N$, resulting in inefficiency ratios greater than $10$ for $N=10$, indicating more than a tenfold loss in social reward. This aligns with \Cref{thm:PoA_selfish}.
Similarly, $\text{IR}^{\emptyset}$ under the hiding mechanism exhibits comparable performance loss to the selfish policy, validating our analysis in \Cref{lemma:information_mechanism}.
In contrast, $\text{IR}^{(\text{CISP})}$ under our CISP mechanism remains around $1$, in line with \Cref{thm:RSP_PoA}. 
In Fig.~\ref{subfig:avg_rho0.95}, with a large discount factor $\rho=0.95$, 
we show that even if the initial belief $\theta_1^n=0.99$, non-myopic players' long-term planning motivates them to explore other arms $k\neq 1$ for their long-term rewards, as per their threshold-based exploration solutions in \Cref{prop:selfish_threshold}. 
Despite of this, both selfish policy and hiding mechanism still result in an increase of $\text{IR}^{(s)}$ and $\text{IR}^{\emptyset}$ to $2$ for $N=10$, due to players selfishly sticking with the best arm 1. However, our CISP mechanism efficiently reduces $\text{IR}^{(\text{CISP})}$ to around $1$, approaching the socially optimal exploration-exploitation performance.

\section{CONCLUSION}\label{section6}
In this paper, we study a new $N$-player competitive MAB game for resource sharing, where non-myopic players do not communicate but compete with each other to form diverse private estimations of unknown arms over time. 
We have explicitly solved both policies in terms of threshold-based structures.
We analyze that in CMAB games, the selfish policy causes the convergence time \(\Omega\big(\frac{K}{\eta^2}\ln({\frac{KN}{\delta}})\big)\) to explore arms, while the socially optimal policy with coordinated communication reduces it to \(\mathcal{O}(\frac{K}{N\eta^2}\ln{(\frac{K}{\delta})})\).
Based on the policy comparison, we prove that the competition among selfish players for the best arm can result in an infinite price of anarchy (PoA), indicating an arbitrarily large efficiency loss compared to the social optimum. 
We further prove that no informational (non-monetary) mechanism (including Bayesian persuasion) can reduce the infinite PoA, as a non-myopic player may strategically misreport arm observations to gain its own long-term benefit. 
Alternatively, we propose a Combined Informational and Side-Payment (CISP) mechanism, which provides socially optimal arm recommendations with proper informational and monetary incentives to players according to their diverse and time-varying private beliefs. Our CISP mechanism keeps ex-post budget balanced for the social planner and ensures truthful reporting from players, thereby achieving the minimum \(\text{PoA}=1\) and the same convergence time as the social optimum.

In the future, we aim to extend CMAB games to more realistic applications.
For instance, in wireless networks, neighboring players with conflicting relationships may cause collisions when accessing the same channel to transmit data.
These conflicting relationships are often unknown to the players, making it necessary to learn the conflict graph while exploring different channels.
This introduces a challenging problem that requires designing an efficient mechanism to balance arm exploration with graph learning effectively.



\balance

\begin{IEEEbiography}[{\includegraphics[width=1in,height=1.25in,clip,keepaspectratio]{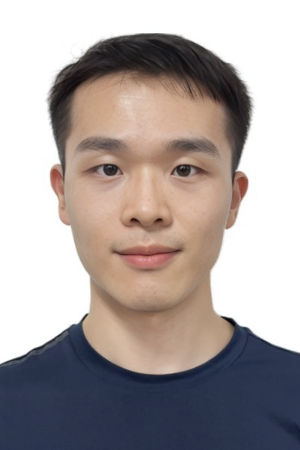}}]{Hongbo Li}(S'24-M'24) received the B.Sc. degree from Shanghai Jiao Tong University (SJTU) in 2019 and the Ph.D. degree from the Singapore University of Technology and Design (SUTD) in 2024. He is currently a Postdoctoral Scholar in the Department of Electrical and Computer Engineering at The Ohio State University (OSU), Columbus, OH, USA. Prior to this, he was a Research Fellow at SUTD in 2024. His research interests include machine learning theory, networked AI, game theory, and mechanism design.
\end{IEEEbiography}

\vspace{11pt}
\begin{IEEEbiography}[{\includegraphics[width=1in,height=1.25in,clip,keepaspectratio]{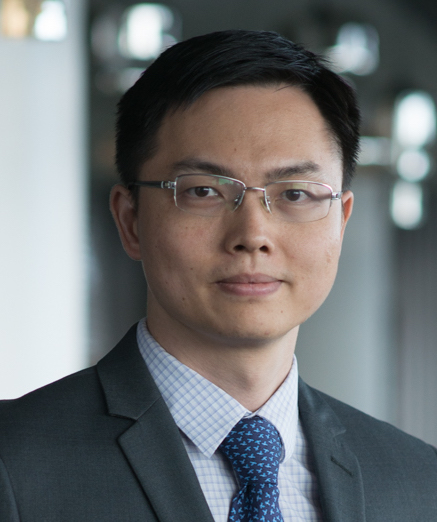}}]{Lingjie Duan}(S'09-M'12-SM'17) received the Ph.D. degree from The Chinese University of Hong Kong in 2012. He is an Associate Professor at the Singapore University of Technology and Design (SUTD) and is an Associate Head of Pillar (AHOP) of Engineering Systems and Design. In 2011, he was a Visiting Scholar at University of California at Berkeley, Berkeley, CA, USA. His research interests include network economics and game theory, network security and privacy, energy harvesting wireless communications, and mobile crowdsourcing. He is an Associate Editor of IEEE/ACM Transactions on Networking and IEEE Transactions on Mobile Computing. He was an Editor of IEEE Transactions on Wireless Communications and IEEE Communications Surveys and Tutorials. He also served as a Guest Editor of the IEEE Journal on Selected Areas in Communications Special Issue on Human-in-the-Loop Mobile Networks, as well as IEEE Wireless Communications Magazine. He is a General Chair of WiOpt 2023 Conference and is a regular TPC member of some other top conferences (e.g., INFOCOM, MobiHoc, SECON). He received the SUTD Excellence in Research Award in 2016 and the 10th IEEE ComSoc Asia-Pacific Outstanding Young Researcher Award in 2015.
\end{IEEEbiography}

\appendix

\subsection{Proof of Lemma 1}
For any player $n$, we suppose it chooses arm $j$ at time $t$. Then its empirical mean reward $\tilde{\mu}^n_i(t+1)$ equals $\tilde{\mu}^n_i(t)$ for any $i\neq j$, as there is no observations on arm $i$. If this user is not selected to pull arm $j$, we obtain $\mathbb{E}[\tilde{\mu}^n_j(t+1)]=\tilde{\mu}^n_j(t)$ for arm $j$.

If it is selected to pull arm $j$, based on eq. (5), we calculate its expected empirical mean reward below:
\begin{align*}
    &\mathbb{E}[\tilde{\mu}^n_j(t+1)]\\=&\mathbb{P}(r_j^n(t)=1)\cdot \frac{\sum_{\tau=1}^t r_j^n(\tau)\mathds{1}\{\pi_n(\tau)=j\}+1}{c_j^n(t)+1}\\&+\mathbb{P}(r_j^n(t)=0)\cdot \frac{\sum_{\tau=1}^t r_k^n(\tau)\mathds{1}\{\pi_n(\tau)=j\}}{c_j^n(t)+1}\\
    =&\tilde{\mu}^n_j(t)\cdot \frac{\sum_{\tau=1}^t r_j^n(\tau)\mathds{1}\{\pi_n(\tau)=j\}+1}{c_j^n(t)+1}\\&+(1-\tilde{\mu}^n_j(t))\cdot \frac{\sum_{\tau=1}^t r_k^n(\tau)\mathds{1}\{\pi_n(\tau)=j\}}{c_j^n(t)+1}\\
    =&\frac{\tilde{\mu}_j^n(t)\cdot(\tilde{\mu}_j^n(t)\cdot c_j^n(t)+1)+(1-\tilde{\mu}_j^n(t))\cdot \tilde{\mu}_j^n(t)\cdot c_j^n(t)}{c_j^n(t)+1}\\
    =&\tilde{\mu}_j^n(t),
\end{align*}
where the second equality is because of $\mathbb{P}(r_j^n(t)=1)=\tilde{\mu}^n_j(t)$ and the third equality is because of $\sum_{\tau=1}^t r_j^n(\tau)\mathds{1}\{\pi_n(\tau)=j\}=\tilde{\mu}_j^n(t)\cdot c_j^n(t)$ derived by eq. (5). Consequently, we obtain $\mathbb{E}[\tilde{\bm{\mu}}^n(t+1)]=\tilde{\bm{\mu}}^n(t)$.

\subsection{Proof of Proposition 1}\label{Appendix_E}
We first prove that when $\rho=0$, player $n$ will switch from arm $k$ to arm $j$ if $\tilde{r}_j^n(t)>\tilde{r}_k^n(t)$, where arm $k$ was the best arm for player $n$ and arm $j$ was the second-best arm for player $n$ at time $t-1$. Then we prove that when $\rho=1$, player $n$ will switch from arm $k$ to arm $j$ if $\tilde{r}_j^n(t)>\tilde{r}_k^n(t)-\Delta \mu_{j,k}$, where $\Delta \mu_{j,k}$ is defined in (10).
Finally, we establish the monotonicity of this exploration threshold $\mathcal{T}^n_{j,k}(t)$, which dictates the minimum immediate reward of arm $j$ for player $n$ to switch to, with respect to $c_j^n(t)$ and $\rho$.

According to the optimization problem (7) under the selfish policy, non-myopic players focus more on long-term rewards as discount factor $\rho\in(0,1)$ increases.
If $\rho=0$, selfish players under (7) become myopic and will switch from arm $k$ to another arm $j$ with a higher immediate expected reward $\tilde{r}_j^n(t)>\tilde{r}_k^n(t)$, regardless of the relationship between $c_j^n(t)$ and $c_k^n(t)$. In this case, the exploration threshold from arm $k$ to arm $j$ satisfies $\mathcal{T}_{j,k}^n(t)=\tilde{r}_k^n(t)$.

If $\rho>0$, player $n$ under selfish policy (7) will explore arm $j$ if it leads to a higher long-term expected reward than arm $k$ from the current time $t$. 
Let $U_t(k,\bm{\pi}^{(s)}_{-n}(t))$ and $U_t(j,\bm{\pi}^{(s)}_{-n}(t))$ denote the long-term expected rewards since time $t$ for player $n$ under current arm decisions $\pi_n(t)=j$ and $\pi_n(t)=k$, respectively.
According to selfish policy (7), player $n$ decides its decision $k$ or $j$ by comparing the two long-term expected rewards below:
\begin{align}
    &U_t\big(k,\bm{\pi}^{(s)}_{-n}(t)\big)-U_t\big(j,\bm{\pi}^{(s)}_{-n}(t)\big)\notag\\=&\tilde{r}_j^n(t)-\tilde{r}_k^n(t)\notag\\&+\rho U_{t+1}\big(\pi_n(t+1|j),\bm{\pi}^{(s)}_{-n}(t+1)\big)\notag\\&-\rho U_{t+1}\big(\pi_n(t+1|k),\bm{\pi}^{(s)}_{-n}(t+1)\big)\notag\\
    =& \frac{\Tilde{\mu}_j^n(t)}{\mathbb{E}\big[|\mathbb{N}_j(t-1)|\big|\tilde{\bm{\mu}}^n(t-1)\big]+1}-\frac{\Tilde{\mu}_k^n(t)}{\mathbb{E}\big[|\mathbb{N}_k(t-1)|\big|\tilde{\bm{\mu}}^n(t-1)\big]}\notag\\&+\rho U_{t+1}\big(\pi_n(t+1|j),\bm{\pi}^{(s)}_{-n}(t+1)\big)\notag\\&-\rho U_{t+1}\big(\pi_n(t+1|k),\bm{\pi}^{(s)}_{-n}(t+1)\big),\tag{21}\label{24}
\end{align}
where $\mathbb{E}\big[|\mathbb{N}_k(t-1)|\big|\tilde{\bm{\mu}}^n(t-1)\big]$ and $\mathbb{E}\big[|\mathbb{N}_j(t-1)|\big|\tilde{\bm{\mu}}^n(t-1)\big]$ are the expected numbers of players choosing arms $k$ and $j$ under $\tilde{\bm{\mu}}^n(t-1)$ at the last time $t-1$, respectively. 

Based on (\ref{24}), the sufficient condition for player $n$ to switch from arm $k$ to arm $j$ is $U_t\big(k,\bm{\pi}^{(s)}_{-n}(t)\big)-U_t\big(j,\bm{\pi}^{(s)}_{-n}(t)\big)<0$, which is equivalent with solving
\begin{align}
    &\rho U_{t+1}\big(\pi_n(t+1|j),\bm{\pi}^{(s)}_{-n}(t+1)\big)\notag\\&-\rho U_{t+1}\big(\pi_n(t+1|k),\bm{\pi}^{(s)}_{-n}(t+1)\big)\tag{22}\label{25}\\>&\frac{\Tilde{\mu}_j^n(t)}{\mathbb{E}\big[|\mathbb{N}_j(t-1)|\big|\tilde{\bm{\mu}}^n(t-1)\big]+1}-\frac{\Tilde{\mu}_k^n(t)}{\mathbb{E}\big[|\mathbb{N}_k(t-1)|\big|\tilde{\bm{\mu}}^n(t-1)\big]}.\notag
\end{align}
To solve the above inequality, we need to expand the two cost-to-go functions $U_{t+1}\big(\pi_n(t+1|j),\bm{\pi}^{(s)}_{-n}(t+1)\big)$ and $U_{t+1}\big(\pi_n(t+1|k),\bm{\pi}^{(s)}_{-n}(t+1)\big)$.
For future arm decisions after choosing arm $\pi_n(i)=j$, player $n$ will switch back to the previously best arm $k$ as long as it pulls arm $j$ and observes $r_j(i)=0$ there. While if it observes $r_j(i)=1$, it will stick with the best arm $j$ at time $i+1$. Therefore, we further expand (\ref{25}) under $\rho=1$ into:
\begin{align}
    & U_{t+1}\big(\pi_n(t+1|j),\bm{\pi}^{(s)}_{-n}(t+1)\big)-U_{t+1}\big(\pi_n(t+1|k),\bm{\pi}^{(s)}_{-n}(t+1)\big)\notag\\ \geq &\Tilde{\mu}_j^n(t) \cdot \frac{\Tilde{\mu}_j^n(t+1|r_j(t)=1)}{\mathbb{E}\big[|\mathbb{N}_j(t-1)|\big|\tilde{\bm{\mu}}^n(t-1)\big]+1}+(1-\Tilde{\mu}_j^n(t))\tilde{r}_k^n(t+1)\notag\\ &-\tilde{r}_k^n(t+1)+\Tilde{\mu}_j^n(t) \cdot\sum_{i=t+2}^{\infty}\big(\tilde{r}_j^n(i)-\tilde{r}_k^n(i)\big)\notag\\
    \geq &\Tilde{\mu}_j^n(t) \frac{1}{\mathbb{E}\big[|\mathbb{N}_j(t-1)|\big|\tilde{\bm{\mu}}^n(t-1)\big]+1}\cdot \frac{\Tilde{\mu}_j^n(t)c_j^n(t)+1}{c_j^n(t)+1}\notag\\&-\Tilde{\mu}_j^n(t)\tilde{r}_k^n(t+1)\tag{23}\label{26}+\Tilde{\mu}_j^n(t) \cdot\sum_{i=t+2}^{\infty}\tilde{r}_j^n(i)-\tilde{r}_k^n(i),
\end{align}
where the second inequality is because of $\Tilde{\mu}_j^n(t+1|r_j(t)=1)=\frac{\Tilde{\mu}_j^n(t)c_j^n(t)+1}{c_j^n(t)+1}$ by (5) and player $n$ switches back to arm $k$ since $t+1$ after observing $r_j(t)=0$ with probability $1-\tilde{\mu}_j^n(t)$.

By combining the above two inequalities in (\ref{25}) and (\ref{26}), we obtain 
\begin{align}
    &\frac{\tilde{\mu}^n_j(t)}{\mathbb{E}\big[|\mathbb{N}_j(t-1)|\big|\tilde{\bm{\mu}}^n(t-1)\big]+1}\tag{24}\label{26-a}\\>&\frac{\tilde{\mu}^n_k(t)}{|\mathbb{N}_k(t-1)|}-\frac{(c_k^n(t)-c_j^n(t))(1-\frac{\tilde{\mu}^n_k(t)}{|\mathbb{N}_k(t-1)|})}{(\mathbb{E}\big[|\mathbb{N}_j(t-1)|\big|\tilde{\bm{\mu}}^n(t-1)\big]+1)(c_k^n(t)c_j^n(t)+c_j^n(t))}.\notag
\end{align}
Letting $\Delta \mu_{j,k}=\frac{|c_k^n(t)-c_j^n(t)|(1-\frac{\tilde{\mu}^n_k(t)}{|\mathbb{N}_k(t-1)|})}{(\mathbb{E}\big[|\mathbb{N}_j(t-1)|\big|\tilde{\bm{\mu}}^n(t-1)\big]+1)(c_k^n(t)c_j^n(t)+c_j^n(t))}$. Then if $c_j^n(t)<c_k^n(t)$, (\ref{26-a}) is the lower bound of threshold $\mathcal{T}_{j,k}^n(t)$ in (9). While if $c_j^n(t)\geq c_k^n(t)$, (\ref{26-a}) becomes the upper bound of $\mathcal{T}_{j,k}^n(t)$ for $c_j^n(t)\geq c_k^n(t)$ in (9). 
Therefore, we derive the exploration threshold $\mathcal{T}_{j,k}^n(t)$ for player $n$ to switch from arm $k$ to arm $j$ when $\rho=1$.

Next, we prove the monotonicity of $\mathcal{T}_{j,k}^n(t)$ with respect to the exploration counts $c_j^n(t)$ and the discount factor $\rho$. By proving $\mathcal{T}_{j,k}^n(t)$ increases with $\rho$ if $c_j^n(t)<c_k^n(t)$ while decreases with $\rho$ if $c_j^n(t)\geq c_k^n(t)$, we can complete the proof of case-by-case ranges of exploration threshold $\mathcal{T}_{j,k}^n(t)$ in (9) of Proposition 1.

In (9), by solving the first-order derivative of $\Delta \mu$ with respect to $c_j^n(t)$, it is straightforward to find that $\Delta \mu$ decreases with $c_j^n(t)$ if $c_j^n(t)<c_k^n(t)$ while increasing with $c_j^n(t)$ if $c_j^n(t)\geq c_k^n(t)$. Thus, $\mathcal{T}_{j,k}^n(t)$ always increases with $c_j^n(t)$, based on the expression in (9). 

To prove $\mathcal{T}_{j,k}^n(t)$ increases/decreases with $\rho$, we only need to show $U_{t+1}\big(\pi_n(t+1|j),\bm{\pi}^{(s)}_{-n}(t+1)\big)-U_{t+1}\big(\pi_n(t+1|k),\bm{\pi}^{(s)}_{-n}(t+1)\big)$ increases/decreases with $\rho$ in (\ref{24}). 

If $c_j^n(t)<c_k^n(t)$, then threshold $\mathcal{T}_{j,k}^n(t)<\frac{\Tilde{\mu}_k^n(t)}{|\mathbb{N}_k(t-1)|}$, as players are always willing to explore arm $j$ with a higher immediate expected reward $\frac{\Tilde{\mu}_j^n(t)}{\mathbb{E}\big[|\mathbb{N}_j(t-1)|\big|\tilde{\bm{\mu}}^n(t-1)\big]+1}$ and less exploration counts $c_j^n(t)$. In this case, 
\begin{align*}
    &\rho U_{t+1}\big(\pi_n(t+1|j),\bm{\pi}^{(s)}_{-n}(t+1)\big)\\&-\rho U_{t+1}\big(\pi_n(t+1|k),\bm{\pi}^{(s)}_{-n}(t+1)\big)>0
\end{align*}
in both (\ref{24}) and (\ref{25}). 
Therefore, a larger discount factor $\rho$ in (\ref{24}) leads to an increased difference between the two cost-to-go, and thus the threshold $\mathcal{T}_{j,k}^n(t)$ also increases. 
While if $c_j^n(t)\geq c_k^n(t)$, we can similarly show
\begin{align*}
    &\rho U_{t+1}\big(\pi_n(t+1|j),\bm{\pi}^{(s)}_{-n}(t+1)\big)\\&-\rho U_{t+1}\big(\pi_n(t+1|k),\bm{\pi}^{(s)}_{-n}(t+1)\big)<0.
\end{align*}
which decreases with $\rho$, indicating that the threshold $\mathcal{T}_{j,k}^n(t)$ decreases with $\rho$.

In summary, If $c_j^n(t)<c_k^n(t)$, this threshold $\mathcal{T}_{j,k}^n(t)$ decreases with discount factor $\rho$. Otherwise, $\mathcal{T}_{j,k}^n(t)$ increases with $\rho$. Based on our derived upper and lower bounds of $\mathcal{T}_{j,k}^n(t)$ under $\rho=1$ and $\rho=0$,
we finally completes the proof of Proposition~1.

\subsection{Proof of Proposition~2}\label{Appendix_A}
We first prove that if any player $n$ has at least $\frac{2}{\eta^2}\ln(\frac{2KN}{\delta})$ observations of each arm, all players will have an $\eta$-correct ranking of all arms with probability at least $1-\delta$. Then we derive the length of exploration $T^{(s)}=\Omega(\frac{K}{\eta^2}\ln(\frac{KN}{\delta}))$ to guarantee the $\frac{2}{\eta^2}\ln(\frac{2KN}{\delta})$ observations for all players. 

Let $O$ denote the required number of observations for each player at each arm. Define $I_n=1$ to be the event that player $n$ does not have an $\eta$-correct ranking. We need to obtain
\begin{align*}
    \mathbb{P}(I_n=1|\text{$\geq O$ observations of each arm})<\frac{\delta}{N}.
\end{align*}
Then we calculate
\begin{align*}
    &\mathbb{P}(I_n=1|\text{$\geq O$ observations of each arm})\\
    \leq & \mathbb{P}\Big(\exists k \text{ s.t. }|\Tilde{\mu}^n_k(t)-\mu_k|>\frac{\eta}{2}\Big|\text{$\geq O$ observations of each arm}\Big)\\
    \leq & \sum_{k=1}^K \mathbb{P}\Big(|\Tilde{\mu}^n_k(t)-\mu_k|>\frac{\eta}{2}\Big|\text{$\geq O$ observations of each arm}\Big)\\
    =& \sum_{k=1}^K\sum_{c=O}^{\infty} \mathbb{P}\Big(|\Tilde{\mu}^n_k(t)-\mu_k|>\frac{\eta}{2}\Big|\# \text{ of observations}=c\Big)\\ &\cdot \mathbb{P}(c \text{ observations}|c\geq O)\\
    \leq &\sum_{k=1}^K\sum_{c=O}^{\infty} 2\exp{(\frac{c\cdot \eta^2}{2})}\mathbb{P}(c \text{ observations}|c\geq O)\\
    \leq &2K\exp{(\frac{-O\cdot \eta^2}{2})},
\end{align*}
where the second inequality is derived by the union bound, the third inequality is derived by Hoeffding's inequality.

To make sure that
\begin{align*}
    \mathbb{P}(I_n=1|\text{$\geq O$ observations of each arm})<\frac{\delta}{N},
\end{align*}
we need $2K\exp{(\frac{-O\cdot \eta^2}{2})}<\frac{\delta}{N}$, solving which we obtain
\begin{align*}
    O>\frac{2}{\eta^2}\ln(\frac{2KN}{\delta}).
\end{align*}

Next, we show that if any player $n$ has at least $\frac{2}{\eta^2}\ln(\frac{2KN}{\delta})$ observations of each arm, all players will have an $\eta$-correct ranking of all arms with probability at least $1-\delta$.  We define the following events:
\begin{itemize}
    \item $L$ denotes the event that all players have an $\eta$-correct ranking.
    \item $L_n$ denotes the event that player $n$ has an $\eta$-correct ranking.
    \item $J$ denotes the event that all players have observed each arm at least $O$ times.
    \item $J_n$ denotes the event that player $n$ has observed each arm at least $O$ times.
\end{itemize}
Then we have 
\begin{align*}
    \mathbb{P}(L|J)&\geq 1-\mathbb{P}(\bigcup_{n\in\mathbb{N}} \Bar{L}_n|J_n )\\ &\geq 1-\sum_{n=1}^N \mathbb{P}(\Bar{L}_n|J_n )\\ &\geq 1-N \frac{\delta}{N}\\&=1-\delta,
\end{align*}
where the second inequality is derived by the union bound.

Finally, we show that there exists a $T^{(s)}$ large enough so that all players have $>O$ observations of each arm with probability at least $1-\delta$. We define $A_{n,k}(t)=1$ to be the event that player $n$ observed arm $k$ at time $t$. By the symmetry property, we have
\begin{align*}
    \mathbb{P}(A_{n,k}(t)=1)=&\mathbb{P}(\pi_{n}^{(s)}(t)=k)\mathbb{P}(\sigma_n(t)=0)\\ \leq& \frac{1}{K},
\end{align*}
where the inequality is because of $\mathbb{P}(\pi_{n}^{(s)}(t)=k)=\frac{1}{K}$ by the symmetry property and $\mathbb{P}(\sigma_n(t)=0)\leq 1$. 
To guarantee that the total number of observations each player has of each arm to be at least $O$, we have
\begin{align*}
    \sum_{t=1}^{T^{(s)}}A_{n,k}(t)\geq T^{(s)}\cdot  \mathbb{E}[A_{n,k}(t)]\geq O,
\end{align*}
solving which we obtain 
\begin{align*}
    T^{(s)}> \frac{2K}{\eta^2}\ln{\left(\frac{2KN}{\delta}\right)}=\Omega\left(\frac{K}{\eta^2}\ln(\frac{KN}{\delta})\right),
\end{align*}
which completes the proof of Proposition~2.

\subsection{Proof of Lemma 1}\label{proof_lemma_nocollision}
We prove Lemma~1 by contradiction. Assume that there exists an arm $k$ that $|\mathbb{N}_k|\geq 2$ under the socially optimal policy, denoted by $\hat{\bm{\pi}}(t)$. We then prove that the long-term total reward for all players under $\hat{\bm{\pi}}(t)$ is smaller than that under $\bm{\pi}^*(t)$ under Lemma~1.

Under $\hat{\bm{\pi}}(t)$, we assume players $n$ and $n'$ choosing the same arm $k$, i.e., $\hat{\pi}_n(t)=\hat{\pi}_{n'}(t)=k$. Under $\bm{\pi}^*(t)$, we assume $\hat{\pi}_n(t)=k$ and $\hat{\pi}_{n'}(t)=k'\neq k$. While for any other player $j\neq n,n'$, we assume $\hat{\pi}_j(t)=\pi^*_j(t)$. Then we only need to compare the summed long-term rewards of players $n$ and $n'$ under the two policies, denoted by $\hat{R}(t)$ and $R^*(t)$, respectively. We calculate:
\begin{align*}
    R^*(t)-\hat{R}(t)=&\tilde{\mu}_k(t)+\tilde{\mu}_{k'}(t)+\rho \mathbb{E}[R^*(t+1)]\\ &-\tilde{\mu}_k(t)-\rho \mathbb{E}[\hat{R}(t+1)],
\end{align*}
where the first equality is because players $n$ and $n'$ receive a total reward of $\tilde{\mu}_k(t)+\tilde{\mu}_{k'}(t)$ under $\bm{\pi}^*(t)$, while they only receive $\tilde{\mu}_k(t)$ under $\hat{\bm{\pi}}(t)$ due to collision. For arm $k'$ that explored by player $n'$ under $\bm{\pi}^*(t)$, its expected empirical mean reward at the next time $t+1$ satisfies
\begin{align*}
    \mathbb{E}[\Tilde{\mu}_k(t+1)]=&\Tilde{\mu}_k(t+1) \mathbb{E}[\Tilde{\mu}_k(t+1)|r_k(t)=1]\\&+(1-\Tilde{\mu}_k(t+1))\mathbb{E}[\Tilde{\mu}_k(t+1)|r_k(t)=0]\\
    =&\Tilde{\mu}_k(t+1) \cdot \frac{\Tilde{\mu}_k(t+1) c_k(t)+1}{c_k(t)+1}\\&+(1-\Tilde{\mu}_k(t+1))\cdot \frac{\Tilde{\mu}_k(t+1) c_k(t)}{c_k(t)+1}\\
    =&\Tilde{\mu}_k(t+1),
\end{align*}
where the second equality is derived based on the calculation of empirical mean reward in (5). Therefore, given $\mathbb{E}[R^*(t+1)]=\mathbb{E}[\hat{R}(t+1)]$ under both policies, we obtain $\mathbb{E}[\hat{R}(t+1)]\leq \mathbb{E}[R^*(t+1)]$, as both policies can always make the same decisions for players $n$ and $n'$. Consequently, we obtain
\begin{align*}
    R^*(t)-\hat{R}(t)&=\tilde{\mu}_{k'}(t)+\rho (\mathbb{E}[R^*(t+1)]-\mathbb{E}[\hat{R}(t+1)])\\&\geq \tilde{\mu}_{k'}(t)>0,
\end{align*}
meaning that policy $\bm{\pi}^*(t)$ with $|\mathbb{N}_k|\leq 1$ outperforms $\hat{\bm{\pi}}(t)$ with $|\mathbb{N}_k|= 2$. This completes the proof of Lemma~1.

\subsection{Proof of Proposition~3}\label{Appendix_F}
We will use the same method as in \Cref{Appendix_E} to prove threshold $\mathcal{T}_{j,k}(t)$ in (12) under socially optimal policy (8).

According to Lemma~1, the number of players choosing any arm $k\in\mathbb{K}$ satisfies $|\mathbb{N}_k(t)|\leq 1$ for any time $t$ under the socially optimal policy. Therefore, the social optimum always chooses $N$ arms to avoid collisions.

If $\rho=0$, the socially optimal policy always maximizes the immediate total reward of all players. In this case, the exploration threshold from arm $k$ to arm $j$ satisfies
\begin{align*}
    \mathcal{T}^*_{j,k}(t)=\tilde{r}_k(t)=\Tilde{\mu}_k(t),
\end{align*}
based on the fact that $\mathbf{Pr}(\sigma_n(t)=0)=1$ for any player $n$ under the socially optimal policy.
 
If $\rho=1$, let $R^*_j$ and $R^*_k$ denote the long-term socially optimal rewards for all players under $\pi^*_n(t)=j$ and $\pi^*_n(t)=k$, respectively. We can calculate 
\begin{align*}
    R^*_j-R^*_k=&\sum_{i=t}^{\infty}(\tilde{r}_j^n(i)-\tilde{r}_k^n(i))\\
    =&\Tilde{\mu}_j(t)-\Tilde{\mu}_k(t)+\sum_{i=t+1}^{\infty}(\tilde{r}_j(i)-\tilde{r}_k(i)).
\end{align*}
Based on this equation, we similarly calculate $\tilde{r}_j(i)-\tilde{r}_k(i)$ as in (\ref{26}) to derive
\begin{align*}
    \tilde{\mu}_j(t)>\tilde{\mu}_k(t)-\frac{(c_k(t)-c_j(t))(1-\tilde{\mu}_k(t))}{c_k(t)c_j(t)+c_j(t)},
\end{align*}
by substituting $|\mathbb{N}_k(t-1)|=1$ and $\mathbb{E}\big[|\mathbb{N}_j(t-1)|\big|\tilde{\bm{\mu}}^n(t-1)\big]=0$ into (\ref{26-a}). Let $\Delta \mu^*_{j,k}=\frac{(c_k(t)-c_j(t))(1-\tilde{\mu}_k(t))}{c_k(t)c_j(t)+c_j(t)}$, we derive the optimal threshold $\mathcal{T}^*_{j,k}(t)$ (12) in Proposition~3.
Additionally, based on the above analysis, $\mathcal{T}^*_{j,k}(t)$ holds the same monotonicity as $\mathcal{T}_{j,k}^n(t)$ in (9) under the selfish policy.

\subsection{Proof of Proposition~4}\label{Appendix_C}
We first prove that if the social planner has at least $\frac{2}{\eta^2}\ln(\frac{2}{\delta})$ observations of each arm, it will have an $\eta$-correct ranking of all arms with probability at least $1-\delta$. Then we derive the length of exploration $T^*=\mathcal{O}(\frac{K}{N\eta^2}\ln(\frac{K}{\delta}))$ to guarantee the $\frac{2}{\eta^2}\ln(\frac{2}{\delta})$ observations for the social planner. 

Let $O$ denote the required number of observations for the social planner at each arm. Define $I=1$ to be the event that the social planner does not have an $\eta$-correct ranking. We need to obtain
\begin{align*}
    \mathbb{P}(I=1|\text{$\geq O$ observations of each arm})<\delta.
\end{align*}

Then we similarly calculate
\begin{align*}
    \mathbb{P}(I=1|\text{$\geq O$ observations of each arm})\leq &2K\exp{(\frac{-O\cdot \eta^2}{2})},
\end{align*}

To make sure that
\begin{align*}
    \mathbb{P}(I=1|\text{$\geq O$ observations of each arm})<\delta,
\end{align*}
we need $2K\exp{(\frac{-O\cdot \eta^2}{2})}<\delta$, solving which we obtain $O>\frac{2}{\eta^2}\ln(\frac{2K}{\delta})$.
Then given $O$ observations of each arm, there is a probability at least $1-\delta$ that the social planner has an $\eta$-correct ranking.

Finally, we show that there exists a $T^*$ large enough so that the social planner has $>O$ observations of each arm with probability at least $1-\delta$. Under Lemma~1, we have the social planner can have $N$ observations at each time $t$, by letting each player choose a specific arm. Therefore, we obtain the probability that the social planner observed arm $k$ at time $t$ is $\frac{N}{K}$. Then we calculate
\begin{align*}
    T^*=\frac{OK}{N}=\mathcal{O}(\frac{K}{N\eta^2}\ln(\frac{K}{\delta}))
\end{align*}
This completes the proof of Proposition~3.

\subsection{Proof of Theorem~1}
We first consider a worst-case scenario with maximum collisions on the best arm (arm 1) and zero exploration of other arms. 
This scenario provides a lower bound, $\text{PoA}\geq 1+\frac{\sum_{k=2}^N\mu_k}{\mu_1}$, which depends on the actual reward $\mu_1\rightarrow 1$ and initial private preference $\theta_1^n>N\theta_k^n$ for any player~$n$ and any arm $k\in\{2,\cdots, N\}$. 
We then demonstrate that this scenario also leads to the upper bound, $\text{PoA}\leq 1+\frac{\sum_{k=2}^N \mu_k}{\mu_1}$. Using these bounds, we derive the closed-form expression of PoA in (15). 

\emph{Lower bound of PoA in (15).} 
Given $\theta_1^n>N\theta_k^n$ and for any arm $k\in\{2,\cdots, K\}$ and any player $n\in\mathbb{N}$, and under the condition $\rho\rightarrow 0$, the exploration threshold of any other arm $k$ satisfies $\mathcal{T}_{k,1}^n(1)=\frac{\theta_1^n}{N}$ as derived in Proposition~1. 
Since $\theta_k^n<\frac{\theta_1^n}{N}=\mathcal{T}_{k,1}^n(1)$, all $N$ selfish players will choose arm 1 at time $t=1$. 
Given $\mu_1\rightarrow 1$, the empirical mean reward $\Tilde{\mu}_1^n(t)=\mu_1\rightarrow 1$ holds true for any player $n$ at any $t$. Thus, each player's actual immediate expected reward is always $\frac{\mu_1}{N}$, derived by (6). 

Let $\underline{R}^{(s)}$ denote the minimum long-term total reward for all players under selfish policy (7) under our setting described above. Based on our analysis, we obtain
\begin{align}\label{27}\tag{25}
    \underline{R}^{(s)}=\sum_{t=1}^{\infty}\sum_{n=1}^N \rho^{t-1} \tilde{r}_1^n(t)=\sum_{t=1}^{\infty}\rho^{t-1} \mu_1=\frac{\mu_1}{1-\rho}.
\end{align}

According to Lemma~1, the socially optimal policy ensures that $N$ players choose arms $1,2,\cdots, N$ starting from time $t=1$.
Given the actual mean reward $\mu_k\rightarrow 1$ for each chosen arm $k\in\{2,\cdots,N\}$, the optimal long-term total expected reward, denoted by $\overline{R}^*$, is 
\begin{align}
    \overline{R}^*&=\sum_{t=1}^{\infty}\sum_{n=1}^N \rho^{t-1} \mathbb{E}[r_{k}^n(t)|\mu_k,\forall k\in\{1,\cdots, N\}]\notag \\ &=\sum_{t=1}^{\infty}\rho^{t-1} (\mu_1+\mu_2+\cdots +\mu_N)\notag\\ \label{28}\tag{26}&=\frac{\mu_1+\sum_{k=2}^{N} \mu_k}{1-\rho}.
\end{align}

As PoA is the maximum ratio between the long-term total rewards under the two policies, we substitute the above two long-term total rewards $\underline{R}^{(s)}$ in (\ref{27}) and $\overline{R}^*$ in (\ref{28}) of this special case into (14) to obtain
\begin{align}\label{29}\tag{27}
    \text{PoA}\geq \frac{\overline{R}^*}{\underline{R}^{(s)}}=1+\frac{\sum_{k=2}^{N}\mu_k}{\mu_1}.
\end{align}

Next, we prove $1+\frac{\sum_{k=2}^{N}\mu_k}{\theta_1}$ in (\ref{29}) is also the upper bound of PoA. 

\emph{Upper bound of PoA in (15).} Under the selfish policy (7), players make their arm decisions in a fully distributed way. 
The worst-case scenario occurs when all the $N$ players stick with the same arm maximum collisions from $t=1$. This results in the minimum long-term reward $R^{(s)}\geq \underline{R}^{(s)}$ in (\ref{27}). 

For the socially optimal policy, it controls the players to choose $N$ different arms to minimize collisions. 
Since the mean actual reward of each arm $k\in\{1,\cdots, K\}$ satisfies $\mu_k< 1$, the total long-term reward reaches a maximum of:
\begin{align*}
    R^*\leq \sum_{t=1}^\infty \rho^{t-1} N =\frac{N}{1-\rho}=\overline{R}^*,
\end{align*}
indicating that the long-term total reward $\overline{R}^*$ in (\ref{28}) is the maximum possible under any conditions. 

Based on the above analysis of the upper and lower bounds, we finally obtain the closed-form expression for PoA in (15):
\begin{align*}
    \text{PoA}=\max \frac{R^*}{R^{(s)}}\leq \frac{\overline{R}^*}{\underline{R}^{(s)}}=1+\frac{\sum_{k=2}^{N}\mu_k}{\mu_1},
\end{align*}
which approaches infinity as $N\rightarrow \infty, K\rightarrow \infty$ and $\mu_k\rightarrow 1$ for any arm $k\in\{2,\cdots, N\}$.

\subsection{Proof of Lemma 2}\label{Appendix_I}

To prove this lemma, we consider the same worst-case scenario with $\theta_1^n>N\theta_k^n$ and for any arm $k\in\{2,\cdots, K\}$ any player $n\in\mathbb{N}$ in Theorem~1. 
Let $I_n(t)$ denote the information incentives that player $n$ received from an informational mechanism at time $t$. 
Then we prove Lemma 2 by mathematical induction.

At initial time $t=1$, given $\theta_1^n>N\theta_k^n$ for any arm $k\in\{2,\cdots,K\}$ and incentive $I_n(1)$, player $n$'s expected empirical mean reward of choosing arm 1 is 
\begin{align*}
    \mathbb{E}[r_1^n(0)|\theta^n_1,I_n(1)]=\theta^n_1.
\end{align*}
According to Proposition~1, player $n$'s exploration threshold of any other arm $k$ satisfies $\mathcal{T}_{k,1}^n(1)=\frac{\theta_1^n}{N}$. As $\theta_k^n<\frac{\theta_1^n}{N}=\mathcal{T}_{k,1}^n(1)$, any player $n\in\mathbb{N}$ will choose arm 1 at time $t=1$. 

Next, we assume that $\pi^{(s)}_n(\tau)=1$ is always true for any $\tau\in\{2,\cdots, t-1\}$. 
Given $\mu_1\rightarrow 1$, according to (5), player $n$'s empirical mean reward satisfies
\begin{align*}
    \tilde{\mu}_1^n(t)=\begin{cases}
        \theta_1^n, &\text{if }c_k^n(t)=0,\\
        \mu_1, &\text{otherwise,}
    \end{cases}
\end{align*}
where $c_k^n(t)$ is the number of times that player $n$ selected to pull arm $k$, as defined in (4). Then for any other arm $k\in\{2,\cdots, N\}$, player $n$'s empirical mean reward is $\tilde{\mu}_k^n(t)=\theta_k^n$, due to the fact that $\pi^{(s)}_n(\tau)\neq k$ for any $\tau<t$. 

Then at time $t$, as $\mathbb{E}[\tilde{\mu}_1^n(t)|I_n(t)]=\tilde{\mu}_1^n(t)>N \tilde{\mu}_k^n(t)$ for any arm $k\in\{2,\cdots, N\}$, player $n$ still chooses arm 1 at time $t$. This completes the proof that player $n$ will not follow any informational mechanism. According to our analysis of Theorem~1, the caused PoA is still infinity.



\subsection{Proof of Proposition 5}
We prove players' long-term incentive compatibility under our four-step CISP mechanism by considering the following three player groups at any time $t$. 
\begin{itemize}
    \item \emph{Players choosing non-optimal arms ($\pi_n^{(s)}(t)\notin K^*(t)$):} For player $n$ choosing arm $\pi_n^{(s)}(t)\notin K^*(t)$ that is not in the set of optimal arms $K^*(t)$, it is persuaded by recommended information $\pi_n^*(t)=k\in K^*(t)$ in step~2. 
    Given the optimal recommendation $\pi_n^*(t)=k$, the player infers that $\tilde{\mu}_k(t)$ exceeds the optimal threshold $\mathcal{T}^*_{k,\pi_n^{(s)}(t)}(t)$ in (9). 
    Since selfish threshold in (9) satisfies $\mathcal{T}^n_{k,\pi_n^{(s)}(t)}(t)\leq \mathcal{T}^*_{k,\pi_n^{(s)}(t)}(t)$, the player further infers \[\tilde{\mu}_k(t)>\mathcal{T}^n_{k,\pi_n^{(s)}(t)}(t).\] Thus, by Proposition~1, following the recommendation $\pi_n^*(t)=k$ is long-term incentive-compatible for player $n$.
    
    \item \emph{Players with maximum reward on optimal arms ($\pi_n^{(s)}(t)=i\in\mathbb{K}^*$ with $|\mathbb{N}_i(t)|>1$):} The player $n=\arg\max_{h\in \mathbb{N}_i(t)}\{\tilde{\mu}_i^h(t)\}$ who has the maximum empirical mean reward $\tilde{\mu}_i^h(t)$ among those choosing arm $i\in \mathbb{K}^*(t)$ is charged $p_i(t)$ in (19) of step 3 to stay on arm $\pi_n^*(t)=\pi_n^{(s)}(t)=i$ in step~3. 
    Without our CISP mechanism, based on the competitive reward model in Section~2.1, each player $n\in\mathbb{N}_i(t)$ choosing arm $\pi_n^{(s)}(t)=i$ under its private information $\tilde{\mu}_i^n(t)$ would receive an expected reward if $\frac{\tilde{\mu}_i^n(t)}{|\mathbb{N}_i(t)|}$. 
    In step 3, if the other $|\mathbb{N}_i(t)|-1$ players follow optimal recommendations to switch to other arms, the charged player $n$ becomes the sole player pulling arm $i$. 
    Thus, with the charged payment $p_i(t)$ in (19), player $n$'s expected reward for choosing arm $i$ is \[\Tilde{\mu}_i^n(t)-p_i(t)=\frac{\Tilde{\mu}_i^n(t)}{|\mathbb{N}_i(t)|},\]
    which does not reduce the original immediate reward without our CISP mechanism.
    As others are persuaded to other arms by rewarding (21) to avoid collisions, staying on arm $\pi_n^{(s)}(t)=i$ for better exploration is long-term incentive-compatible for this player $n$.
    
    \item \emph{Players rewarded to change arms ($\pi_l^{(s)}(t)=i$ to $\pi^*_l(t)=j$):} For player $l$ satisfying (20), it is rewarded $p_j(t)$ in (21) to change from the selfish arm $\pi_l^{(s)}(t)=i$ to the optimal arm $\pi_l^*(t)=j$ in step~3. 
    Without the CISP mechanism, player $l$ would choose arm $i$ to receive an expected reward of $\frac{\tilde{\mu}_i^l(t)}{|\mathbb{N}_i(t)|}$. 
    Under our CISP mechanism, assuming other players follow our CISP and change their arm decisions, player $l$'s immediate reward for choosing arm $\pi^*_l(t)$ becomes
    \[p_j(t)+\tilde{\mu}^l_j(t)=\frac{\tilde{\mu}_i^n(t)}{|\mathbb{N}_i(t)|},\]
    which is higher than its original reward $\frac{\tilde{\mu}_i^l(t)}{|\mathbb{N}_i(t)|}$, where $n=\arg\max_h\{\tilde{\mu}_{\pi_l^{(s)}(t)}^h(t)\}$ in (21). 
    Therefore, player $l$'s immediate reward increases by following arm recommendation $\pi^*_l(t)=j$. 
    Note that player $l$ may consider deviating back to arm $i$, anticipating it would be the sole player under the CISP mechanism. 
    If it deviates to arm $i$, its immediate reward becomes $\frac{\tilde{\mu}_i^l(t)}{2}$, which may be higher than $\frac{\tilde{\mu}_i^n(t)}{|\mathbb{N}_i(t)|}$ of arm $\pi_l^*(t)=j$. 
    However, if it deviates, there will be a penalty in step 4, which makes its immediate reward much lower than $\frac{\tilde{\mu}_i^n(t)}{|\mathbb{N}_i(t)|}$. Therefore, deviating to arm $\pi^{(s)}(t)=i$ results in a lower reward than following arm $\pi^*(t)=j$.
    
    Next, we prove that player $l$’s future reward reduces if it deviates from optimal recommendation $\pi^*(t)=j$ to its selfish arm decision $\pi_l^{(s)}(t)=i$. Notably, it won’t deviate to other arms $k\neq i,j$ without reward. 
    \begin{itemize}
        \item Given others follow CISP’s recommending arms, if player $l$ follows $\pi_l^*(t)=j$, its empirical rewards for other arms remain. The player's selfish policy (7) at any $\tau\in\{t+1,\cdots\}$ has two possibilities: $\pi_l^{(s)}(\tau)=i$ if low-reward observed at arm $j$ or $\pi_l^{(s)}(\tau)=j$ if high-reward at $j$. In the first case, player $l$ consistently receives higher immediate rewards at any $\tau\in\{t+1,\cdots\}$ by receiving reward (21) from our CISP mechanism. In the second case, though without reward (21), arm $\pi_l^*(\tau)=\pi_l^{(s)}(\tau)=j$ has a higher empirical mean reward than $\frac{\tilde{\mu}_i^n(\tau)}{|\mathbb{N}_i(\tau)|}$ to attract it. Denote by $R^*_j(t)$ the total future reward since $t+1$ for player $n$ by choosing arm $\pi^*(t)=j$ at current time $t$. In summary, its future reward of following $\pi_l^*(t)=j$ satisfies
        \[R^*_j(t)\geq \sum_{\tau=t+1}\rho^{\tau-t}\frac{\tilde{\mu}_i^n(\tau)}{|\mathbb{N}_i(\tau)|}.\]
        \item If player $l$ deviates to arm $\pi_l^{(s)}(t)=i$ at current time $t$, even with high future empirical reward $\tilde{\mu}_i^l(\tau)$, it will be charged (19) from $t+1$ to receive immediate rewards no higher than $\frac{\tilde{\mu}_i^l(\tau)}{|N_i(\tau)|}$. 
        Let $R_i^{(s)}(t)$ denote player $n$'s total future reward from $t+1$ of deviating to arm $\pi^{(s)}(t)=i$. Considering (5), the future reward (even without collisions) for deviating to $\pi_l^{(s)}(t)=i$ satisfies
        \begin{align*}
            R_i^{(s)}(t)&\leq \sum_{\tau=t+1}\rho^{\tau-t}\frac{\tilde{\mu}_i^l(\tau)}{|\mathbb{N}_i(\tau)|}\\ &\leq\sum_{\tau=t+1}\rho^{\tau-t}\frac{\tilde{\mu}_i^n(\tau)}{|\mathbb{N}_i(\tau)|} \\&\leq R_j^*(t),
        \end{align*}
        indicating that the future total reward for deviating to arm $\pi_l^{(s)}(t)=i$ will never exceed that of following arm $\pi_l^*(t)=j$ under our CISP mechanism.
    \end{itemize}
    Based on this analysis, player $l$ achieves both higher immediate and future total rewards by choosing arm $\pi_l^*(t)=j$ rather than deviating to arm $\pi_l^{(s)}(t)=j$. Therefore, following the CISP's optimal recommendation $\pi^*_l(t)=j$ is long-term incentive compatible for player $l$.
\end{itemize}

As a result, at any time $t$, all players can expect a long-term total reward at least as good as their outside options, making them long-term incentive-compatible to participate in our CISP mechanism. 

Finally, we prove that all players are individually rational to always follow our CISP mechanism. 
\begin{itemize}
    \item On one hand, according to our above proof of incentive compatibility, our CISP mechanism allows each player $n$ to adhere to its optimal arm $\pi_n^*(t)\in\mathbb{K}^*(t)$ from the initial time $t=0$, where $\mathbb{K}^*(t)$ is the optimal arm set. Assume other players follow our CISP mechanism's recommendations. If player $n$'s selfish arm decision $\pi_n^{(s)}(t)\in \mathbb{K}^*(t)$ is in the set of optimal arms but different from the optimal recommendation $\pi_n^{(s)}(t)\neq \pi_n^*(t)$, the player will not deviate to another arm $k\in \mathbb{K}^*(t)$, where $k\neq \pi_n^*(t)$. This is because deviation would result in a collision with another player on arm $k$, leading to a lower reward and a penalty $\Psi\gg 1$ in step 4 of our CISP mechanism. Thus, the player will continue choosing arm $\pi^*_n(t)$ at any time $t$, and misreporting its observation will not enhance its expected reward on this arm. Therefore, our CISP mechanism keeps individual rationality for player $n$.
    
    \item On the other hand, if player $n$'s selfish arm decision s not in the optimal set $\pi^{(s)}_n(t)\notin \mathbb{K}^*(t)$, it may consider misreporting its observation on arm $\pi^*_n(t)$ to deviate to arm $\pi^{(s)}_n(t)$. However, $\pi^{(s)}_n(t)\notin \mathbb{K}^*(t)$ cannot be true, because each player's selfish exploration threshold in (9) satisfies $\mathcal{T}^n_{k,\pi_n^{(s)}(t)}(t)\leq \mathcal{T}^*_{k,\pi_n^{(s)}(t)}(t)$, where $\mathcal{T}^*_{k,\pi_n^{(s)}(t)}(t)$ is the optimal exploration threshold in (12). Therefore, any selfish player $n$ will not deviate to another arm $k\notin \mathbb{K}^*(t)$. 
\end{itemize}
In conclusion, all players are individually rational to stick with the recommended arm $\pi^*(t)$ and truthfully report their observed rewards. This completes the proof of Proposition~5.

\subsection{Proof of Lemma 3}\label{Appendix_proof_lemmaBB}

According to the side-payment incentives in step 3 of Definition~7, our CISP mechanism charges a payment $p_i(t)$ in (19) from player $l\in\mathbb{N}_i(t)$ with the maximum empirical mean reward $\tilde{\mu}_i^n(t)$ for arm $i$. Let $\mathbb{N}_i^{-n}$ denote the set of other $|\mathbb{N}_i(t)|-1$ players choosing arm $i$ except for player $n$ being charged. 

For any player $l\in\mathbb{N}_i^{-n}$, its empirical mean reward of arm $i$ satisfies $\tilde{\mu}_i^l<\tilde{\mu}_i^n(t)$.
The social planner rewards $p_i(t)$ in (21) to any player $l\in\mathbb{N}_i^{-n}$ to incentivize a change in its arm decision, and the total reward payment for these $|\mathbb{N}_i(t)|-1$ players is given by
\begin{align*}
    \sum_{l=1}^{|\mathbb{N}_i(t)|-1}p_j(t)&= \sum_{l=1}^{|\mathbb{N}_i(t)|-1} \left(\frac{\tilde{\mu}_{i}^n(t)}{|\mathbb{N}_{i}(t)|}-\Tilde{\mu}_j^l(t)\right)\\
    &<\sum_{l=1}^{|\mathbb{N}_i(t)|-1} \frac{\tilde{\mu}_{i}^n(t)}{|\mathbb{N}_{i}(t)|}\\
    &= \frac{|\mathbb{N}_{i}(t)|-1}{|\mathbb{N}_{i}(t)|}\tilde{\mu}_{i}^n(t),
\end{align*}
where arm $j$ is derived by (20) as player $l$'s optimal arm recommendation $\pi^*_l(t)=j$ in Definition~7. 
Thus, the charged amount $p_i(t)$ is sufficient to compensate the other $|\mathbb{N}_i(t)|-1$ players for switching to other arms $j\in\mathbb{K}^*(t)$ with $|\mathbb{N}_j(t)|=0$. 
In other words, the social planner maintains ex-post budget balanced at any time.

\subsection{Proof of Theorem 2}\label{Appendix_K}
Based on the proof of Proposition~5, players are long-term incentive-compatible and individually rational, always reporting truthfully and following the optimal recommendations under our CISP mechanism. 
Consequently, our CISP mechanism successfully changes the sub-optimal player distributions from $|\mathbb{N}_i(t)|>1$ and $|\mathbb{N}_j(t)|=0$ to $|\mathbb{N}_i(t)|=1$ and $|\mathbb{N}_j(t)|=1$ for any arm $i,j \in \mathbb{K}^*(t)$ at all time. 
Under these optimal arm choices for any time $t$, the resulting PoA by our CISP mechanism is the optimum, $\text{PoA}=1$.
The convergence period also becomes $\mathcal{O}\big(\frac{K}{N\eta^2}\ln{(\frac{K}{\delta})}\big)$, identical to that of the socially optimal policy as stated in Proposition~3.


\begin{thebibliography}{10}
\providecommand{\url}[1]{#1}
\csname url@samestyle\endcsname
\providecommand{\newblock}{\relax}
\providecommand{\bibinfo}[2]{#2}
\providecommand{\BIBentrySTDinterwordspacing}{\spaceskip=0pt\relax}
\providecommand{\BIBentryALTinterwordstretchfactor}{4}
\providecommand{\BIBentryALTinterwordspacing}{\spaceskip=\fontdimen2\font plus
\BIBentryALTinterwordstretchfactor\fontdimen3\font minus
  \fontdimen4\font\relax}
\providecommand{\BIBforeignlanguage}[2]{{%
\expandafter\ifx\csname l@#1\endcsname\relax
\typeout{** WARNING: IEEEtran.bst: No hyphenation pattern has been}%
\typeout{** loaded for the language `#1'. Using the pattern for}%
\typeout{** the default language instead.}%
\else
\language=\csname l@#1\endcsname
\fi
#2}}
\providecommand{\BIBdecl}{\relax}
\BIBdecl

\bibitem{li2020multi}
F.~Li, D.~Yu, H.~Yang, J.~Yu, H.~Karl, and X.~Cheng, ``Multi-armed-bandit-based
  spectrum scheduling algorithms in wireless networks: A survey,'' \emph{IEEE
  Wireless Communications}, vol.~27, no.~1, pp. 24--30, 2020.

\bibitem{jiang2025multi}
Y.~Jiang, J.~Ye, L.~Zhou, X.~Ge, J.~Kang and D.~Niyato, ``Multi-Modal Stream Integrity Transmission Strategy for Multi-User Wireless Metaverse,'' \emph{IEEE Transactions on Communications}, 2025.

\bibitem{liu2025optimizing}
J.~Liu, M.~Xiao, J.~Wen, J.~Kang, R.~Zhang, T.~Zhang, D.~Niyato, W.~Zhang, and Y.~Liu, ``Optimizing Resource Allocation for Multi-Modal Semantic Communication in Mobile AIGC Networks: A Diffusion-Based Game Approach,'' \emph{IEEE Transactions on Cognitive Communications and Networking}, 2025.

\bibitem{liu2018information}
F.~Liu, S.~Buccapatnam, and N.~Shroff, ``Information directed sampling for
  stochastic bandits with graph feedback,'' in \emph{Proceedings of the AAAI
  Conference on Artificial Intelligence}, vol.~32, no.~1, 2018.

\bibitem{yang2022distributed}
L.~Yang, Y.-Z.~J. Chen, M.~H. Hajiemaili, J.~C. Lui, and D.~Towsley,
  ``Distributed bandits with heterogeneous agents,'' in \emph{IEEE INFOCOM 2022-IEEE Conference on Computer Communications},\hskip 1em plus 0.5em minus
  0.4em\relax IEEE, pp. 200--209, 2022.

\bibitem{zhu2023distributed}
J.~Zhu and J.~Liu, ``Distributed multi-armed bandits,'' \emph{IEEE Transactions
  on Automatic Control}, vol.~68, no.~5, pp. 3025--3040, 2023.

\bibitem{li2023congestion}
H.~Li and L.~Duan, ``When congestion games meet mobile crowdsourcing: Selective
  information disclosure,'' in \emph{Proceedings of the AAAI Conference on
  Artificial Intelligence}, vol.~37, no.~5, pp. 5739--5746, 2023.

\bibitem{li2024human}
H.~Li and L.~Duan, ``Human-in-the-loop learning for dynamic congestion games,'' \emph{IEEE
  Transactions on Mobile Computing}, vol.~23, no.~12, pp. 11159 - 11171, 2024.

  \bibitem{li2024distributed}
H.~Li and L.~Duan, ``Distributed Learning for Dynamic Congestion Games,'' in \emph{IEEE International Symposium on Information Theory (ISIT)}, pp. 3654-3659, 2024.

\bibitem{liu2010distributed}
K.~Liu and Q.~Zhao, ``Distributed learning in multi-armed bandit with multiple
  players,'' \emph{IEEE Transactions on Signal Processing}, vol.~58, no.~11,
  pp. 5667--5681, 2010.

\bibitem{marden2014achieving}
J.~R. Marden, H.~P. Young, and L.~Y. Pao, ``Achieving pareto optimality through
  distributed learning,'' \emph{SIAM Journal on Control and Optimization},
  vol.~52, no.~5, pp. 2753--2770, 2014.


\bibitem{rosenski2016multi}
J.~Rosenski, O.~Shamir, and L.~Szlak, ``Multi-player bandits--a musical chairs
  approach,'' in \emph{International Conference on Machine Learning}.\hskip 1em
  plus 0.5em minus 0.4em\relax PMLR, pp. 155--163, 2016.

\bibitem{bistritz2018distributed}
I.~Bistritz and A.~Leshem, ``Distributed multi-player bandits-a game of thrones
  approach,'' \emph{Advances in Neural Information Processing Systems},
  vol.~31, 2018.

\bibitem{boursier2019sic}
E.~Boursier and V.~Perchet, ``Sic-mmab: Synchronisation involves communication
  in multiplayer multi-armed bandits,'' \emph{Advances in Neural Information
  Processing Systems}, vol.~32, 2019.

\bibitem{wang2020optimal}
P.-A. Wang, A.~Proutiere, K.~Ariu, Y.~Jedra, and A.~Russo, ``Optimal algorithms
  for multiplayer multi-armed bandits,'' in \emph{International Conference on
  Artificial Intelligence and Statistics}.\hskip 1em plus 0.5em minus
  0.4em\relax PMLR, pp. 4120--4129, 2020.

\bibitem{xiong2023decentralized}
G.~Xiong and J.~Li, ``Decentralized stochastic multi-player multi-armed walking
  bandits,'' in \emph{Proceedings of the AAAI Conference on Artificial
  Intelligence}, vol.~37, no.~9, pp. 10\,528--10\,536, 2023.

\bibitem{bolton1999strategic}
P.~Bolton and C.~Harris, ``Strategic experimentation,'' \emph{Econometrica},
  vol.~67, no.~2, pp. 349--374, 1999.

\bibitem{branzei2021multiplayer}
S.~Br{\^a}nzei and Y.~Peres, ``Multiplayer bandit learning, from competition to
  cooperation,'' in \emph{Conference on Learning Theory}.\hskip 1em plus 0.5em
  minus 0.4em\relax PMLR, pp. 679--723, 2021.

\bibitem{sentenac2021decentralized}
F.~Sentenac, E.~Boursier, and V.~Perchet, ``Decentralized learning in online
  queuing systems,'' \emph{Advances in Neural Information Processing Systems},
  vol.~34, pp. 18\,501--18\,512, 2021.


\bibitem{boursier2020selfish}
E.~Boursier and V.~Perchet, ``Selfish robustness and equilibria in multi-player bandits,'' in
  \emph{Conference on Learning Theory}.\hskip 1em plus 0.5em minus 0.4em\relax
  PMLR, pp. 530--581, 2020.
  
\bibitem{huang2023near}
Z.~Huang and J.~Pan, ``A near-optimal high-probability swap-regret upper bound
  for multi-agent bandits in unknown general-sum games,'' in \emph{Uncertainty
  in Artificial Intelligence}.\hskip 1em plus 0.5em minus 0.4em\relax PMLR, pp. 911--921,
  2023.
  
\bibitem{xu2023competing}
R.~Xu, H.~Wang, X.~Zhang, B.~Li, and P.~Cui, ``Competing for shareable arms in
  multi-player multi-armed bandits,'' in \emph{International Conference on
  Machine Learning}.\hskip 1em plus 0.5em minus 0.4em\relax PMLR, pp.
  38\,674--38\,706, 2023.
  
\bibitem{kremer2014implementing}
I.~Kremer, Y.~Mansour, and M.~Perry, ``Implementing the “wisdom of the
  crowd”,'' \emph{Journal of Political Economy}, vol. 122, no.~5, pp.
  988--1012, 2014.

\bibitem{papanastasiou2018crowdsourcing}
Y.~Papanastasiou, K.~Bimpikis, and N.~Savva, ``Crowdsourcing exploration,''
  \emph{Management Science}, vol.~64, no.~4, pp. 1727--1746, 2018.

\bibitem{mansour2022bayesian}
Y.~Mansour, A.~Slivkins, V.~Syrgkanis, and Z.~S. Wu, ``Bayesian exploration:
  Incentivizing exploration in bayesian games,'' \emph{Operations Research},
  vol.~70, no.~2, pp. 1105--1127, 2022.

\bibitem{simchowitz2023exploration}
M.~Simchowitz and A.~Slivkins, ``Exploration and incentives in reinforcement
  learning,'' \emph{Operations Research}, 2023.

\bibitem{tavafoghi2017informational}
H.~Tavafoghi and D.~Teneketzis, ``Informational incentives for congestion
  games,'' in \emph{2017 55th Annual Allerton Conference on Communication,
  Control, and Computing (Allerton)}.\hskip 1em plus 0.5em minus 0.4em\relax
  IEEE, pp. 1285--1292, 2017.

\bibitem{das2017reducing}
S.~Das, E.~Kamenica, and R.~Mirka, ``Reducing congestion through information
  design,'' in \emph{2017 55th Annual Allerton Conference on Communication,
  Control, and Computing (Allerton)}.\hskip 1em plus 0.5em minus 0.4em\relax
  IEEE, pp. 1279--1284, 2017.


\bibitem{frazier2014incentivizing}
P.~Frazier, D.~Kempe, J.~Kleinberg, and R.~Kleinberg, ``Incentivizing
  exploration,'' in \emph{Proceedings of the Fifteenth ACM Conference on
  Economics and Computation}, pp. 5--22, 2014.


\bibitem{che2018recommender}
Y.-K. Che and J.~H{\"o}rner, ``Recommender systems as mechanisms for social
  learning,'' \emph{The Quarterly Journal of Economics}, vol. 133, no.~2, pp.
  871--925, 2018.


\bibitem{besson2018multi}
L.~Besson and E.~Kaufmann, ``Multi-player bandits revisited,'' in
  \emph{Algorithmic Learning Theory}.\hskip 1em plus 0.5em minus 0.4em\relax
  PMLR, pp. 56--92, 2018.

\bibitem{krishnasamy2021learning}
S.~Krishnasamy, R.~Sen, R.~Johari, and S.~Shakkottai, ``Learning unknown
  service rates in queues: A multiarmed bandit approach,'' \emph{Operations
  Research}, vol.~69, no.~1, pp. 315--330, 2021.

\bibitem{roughgarden2010algorithmic}
T.~Roughgarden, ``Algorithmic game theory,'' \emph{Communications of the ACM},
  vol.~53, no.~7, pp. 78--86, 2010.


\bibitem{koutsoupias1999worst}
E.~Koutsoupias and C.~Papadimitriou, ``Worst-case equilibria,'' in \emph{Annual
  Symposium on Theoretical Aspects of Computer Science}.\hskip 1em plus 0.5em
  minus 0.4em\relax Springer, pp. 404--413, 1999.


\bibitem{borgers2015introduction}
T.~B{\"o}rgers, \emph{An introduction to the theory of mechanism design}.\hskip
  1em plus 0.5em minus 0.4em\relax Oxford University Press, USA, 2015.

  \bibitem{li2024optimize}
H.~Li and L.~Duan, ``To Optimize Human-in-the-loop Learning in Repeated Routing Games,'' \emph{IEEE
  Transactions on Mobile Computing}, 2024.

  \bibitem{li2025analyze}
H.~Li and L.~Duan, ``To Analyze and Regulate Human-in-the-Loop Learning for Congestion Games,'' in \emph{IEEE Transactions on Networking}, 2025.

\bibitem{mansour2020bayesian}
Y.~Mansour, A.~Slivkins, and V.~Syrgkanis, ``Bayesian incentive-compatible
  bandit exploration,'' \emph{Operations Research}, vol.~68, no.~4, pp.
  1132--1161, 2020.

\bibitem{myerson1979incentive}
R. B.~Myerson, ``Incentive compatibility and the bargaining problem,'' \emph{Econometrica: journal of the Econometric Society}, pp. 61-73, 1979.

\bibitem{gode1993allocative}
D. K.~Gode and S.~Sunder, ``Allocative efficiency of markets with zero-intelligence traders: Market as a partial substitute for individual rationality,'' \emph{Journal of political economy}, vol.~101, no.~1, pp.~119--137, 1993.

\bibitem{wu2019learning}
M.~Wu and S.~Amin, ``Learning an unknown network state in routing games,''
  \emph{IFAC-PapersOnLine}, vol.~52, no.~20, pp. 345--350, 2019.

\bibitem{macault2022social}
E.~Macault, M.~Scarsini, and T.~Tomala, ``Social learning in nonatomic routing
  games,'' \emph{Games and Economic Behavior}, vol. 132, pp. 221--233, 2022.

\bibitem{hayvaci2014spectrum}
H.~Hayvaci, and B.~Tavli, ``Spectrum sharing in radar and wireless communication systems: A review,'' \emph{International Conference on Electromagnetics in Advanced Applications (ICEAA)}, pp. 810--813, 2014.

\bibitem{balakumar2023enhance}
D. Balakumar and N.~Sendrayan, ``Enhance the Probability of Detection of Cooperative Spectrum Sensing in Cognitive Radio Networks Using Blockchain Technology,'' \emph{Journal of Electrical and Computer Engineering}, vol.~2023, no.~1, pp.~8920243, 2023.
  
\end{thebibliography}
\end{document}